\documentclass[aps,prd,nofootinbib, twocolumn,superscriptaddress,preprintnumbers]{revtex4}

\usepackage{amssymb}
\usepackage{amsmath}
\usepackage{epsfig}
\usepackage{hyperref}
\usepackage{breakurl,comment}

\newcommand{\be}{\begin{equation}}
\newcommand{\ee}{\end{equation}}
\newcommand{\bea}{\begin{eqnarray}}
\newcommand{\eea}{\end{eqnarray}}

\newlength {\leftpanesize}
\newlength {\rightpanesize}
\newlength {\titleskip}
\newlength {\startskip}
\newlength {\itemskip}
\newlength {\panesep}
\newlength {\itemparskip}
\newlength {\locationskip}
\newlength {\timeskip}
\newlength {\languagesize}

\begin{document}
\title{Anthropic Origin of the Neutrino Mass from Cooling Failure}
\author{Raphael Bousso}
\author{Dan Mainemer Katz}

\affiliation{Center for Theoretical Physics and Department of Physics, University of California, Berkeley, CA 94720, U.S.A}
\affiliation{Lawrence Berkeley National Laboratory, Berkeley, CA 94720, U.S.A}

\author{Claire Zukowski}

\affiliation{Center for Theoretical Physics and Department of Physics, University of California, Berkeley, CA 94720, U.S.A}
\affiliation{Lawrence Berkeley National Laboratory, Berkeley, CA 94720, U.S.A}
\affiliation{Perimeter Institute for Theoretical Physics, Waterloo, Ontario N2L 2Y5, Canada}

\begin{abstract}
The sum of active neutrino masses is well constrained,  $58$\,meV $\leq m_\nu \lesssim 0.23$\,eV, but the origin of this scale is not well understood. Here we investigate the possibility that it arises by environmental selection in a large landscape of vacua. Earlier work had noted the detrimental effects of neutrinos on large scale structure. However, using Boltzmann codes to compute the smoothed density contrast on Mpc scales, we find that dark matter halos form abundantly for $m_\nu \gtrsim 10$\,eV. This finding rules out an anthropic origin of $m_\nu$, unless a different catastrophic boundary can be identified. Here we argue that galaxy formation becomes inefficient for $m_\nu \gtrsim 10$\,eV. We show that in this regime, structure forms late and is dominated by cluster scales, as in a top-down scenario. This is catastrophic: baryonic gas will cool too slowly to form stars in an abundance comparable to our universe. With this novel cooling boundary, we find that the anthropic prediction for $m_\nu$ agrees at better than $2\sigma$ with current observational bounds. A degenerate hierarchy is mildly preferred.
\end{abstract}

\maketitle

%\tableofcontents

\section{Introduction}
\label{sec-intro}

In a theory with a large multidimensional potential landscape~\cite{BP}, the smallness of the cosmological constant can be anthropically explained~\cite{Wei87}.\footnote{It cannot be explained in a one-dimensional landscape, no matter how large~\cite{Abb85,BroTei87,BroTei88}, because an empty univere is produced. The string theory landscape~\cite{BP,KKLT,DenDou04b} is an example of a multidimensional landscape in which the cosmological constant scans densely and our vacuum can be produced with sufficient free energy. Related early work includes~\cite{CarRee79,DavUnw82,Sak84,Lin84c,Ban85,BarTip86,Lin86a,Lerche:1986cx,Lin87,Efs95,%
Vil95b,TegRee97,Schellekens:2006xz,GarLiv00}. Reviews with varying ranges of detail and technicality are available, for example~\cite{Wei89,Pol06,Bou06b,Bou07,Schellekens:2008kg,Bou12,Schellekens:2013bpa}.} The lack of a viable alternative explanation for a small or vanishing cosmological constant, the increasing evidence for a fine-tuned weak scale, and several other complexity-favoring coincidences and tunings in cosmology and the Standard Model, all motivate us to consider landscape models seriously, and to extract further pre- or post-dictions from them.\looseness=-1

A large landscape can also explain an aspect of the Standard Model that has long remained mysterious: the origin of the masses and mixing angles of the quarks and leptons. Plausible landscape models allow for some of the first generation quark and lepton masses to be anthropically determined, while the remaining parameters are set purely by the statistical distribution of the Yukawa matrices. Results are consistent with the observed hierarchical, generation, and pairing structures~\cite{Don97,Hog00,DonDut05,Damour:2007uv,Donoghue:2009me,HalSal07a,HalSal07b,HalSal08}. In such analyses, the overall mass scale of neutrinos may be held fixed and ascribed, e.g., to a seesaw mechanism. But ultimately, one expects that the mass scale will vary, no matter what the dominant origin of neutrino masses is in the landscape. For Dirac neutrinos, Yukawa couplings can vary; in the seesaw, a coupling or the right-handed neutrino mass scale can vary.

Thus we may ask whether anthropic constraints play a role in determining the overall scale, or sum, of the standard model neutrino masses,
\begin{equation}
m_\nu \equiv m(\nu_e)+m(\nu_\mu)+m(\nu_\tau)~.
\end{equation}
Current observational bounds imply
\begin{equation}
58\,\text{meV}\leq m_\nu \lesssim 0.23\,\text{eV}~.
\label{eq-cb}
\end{equation}
The lower bound comes from the mass splittings observed via solar and atmospheric neutrino oscillations~\cite{Oli14}. The upper bound comes from cosmological observations that have excluded the effects that more massive neutrinos would have had on the cosmic microwave background and on large scale structure~\cite{Hin12,Ade15}. The proximity of the lower to the upper bound gives us confidence that cosmological experiments in the coming decade will detect $m_\nu$, and that they may determine its value with a precision approaching the $10^{-2}$\,eV level~\cite{AbaArn13}.
\begin{figure*}[t]
\centering
\includegraphics[width=6.5 in]{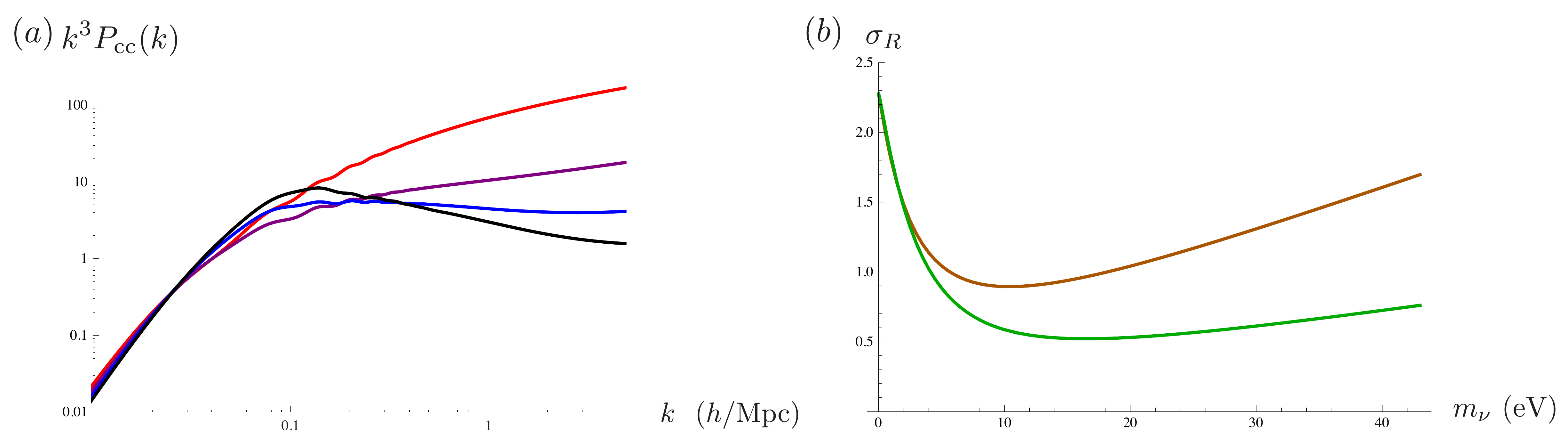}
\caption{(a) The dimensionless power spectrum at $z=0$ for a range of neutrino masses with a normal hierarchy, computed using CAMB. From top to bottom at high wavenumber: $m_\nu = 0$ (red), $m_\nu = 5\mbox{ eV}$ (purple),  $m_\nu = 10\mbox{ eV}$ (blue), and  $m_\nu = 15\mbox{ eV}$ (black). Free-streaming of massive neutrinos causes a suppression of power at high wavenumbers. Above a critical neutrino mass $m \sim 8 \mbox{ eV}$, this effect is large enough so that the dimensionless power spectrum develops a peak near the free-streaming scale $k_{\rm nr}<k_{\rm gal}$. This implies that the first structure consists of cluster-size halos. (b) We obtain the smoothed density contrast $\sigma_R$ at $z=0$ numerically, essentially by integrating $k^3 P(k)$ up to the wavenumber $1/R$; see Eq.~(\ref{eq-srintro}) and surrounding discussion. We take $R$ to be the comoving scale of the Milky Way ($10^{12} M_\odot$). The orange (upper) curve corresponds to a normal hierarchy; the green (lower) to a degenerate hierarchy. We see that neutrinos suppress halo formation only in the regime $m_\nu \lesssim 10$\,eV where the dimensionless power spectrum has no maximum and the integral is dominated by the large-$k$ cutoff. For larger neutrino masses, the formation of galactic and larger halos is actually enhanced, because the dimensionless power spectrum develops a peak that dominates the integral. At higher neutrino masses, the peak power increases, due to a decrease of the free-streaming scale and a lengthening of the matter era. (This is more pronounced for a normal hierarchy.) Hence $\sigma_R$ increases. If observers formed in proportion to the mass fraction in large dark matter halos, this would rule out an anthropic origin of $m_\nu$; see Fig.~\ref{fig-ProbNoCutoff}.}
\label{fig-sigmapower}
\end{figure*}

An anthropic origin of the neutrino mass scale is suggested by the remarkable coincidence that neutrinos have affected cosmology just enough for their effects to be noticeable, but not enough to significantly diminish the abundance of galaxies. A priori, $m_\nu$ could range over dozens of orders of magnitude. If $m_\nu$ was only two orders of magnitude smaller than the observed value, its effects on cosmology would be hard to discern at all. If $m_\nu$ was slightly larger, fewer galaxies would form, and hence fewer observers like us. The goal of this paper is to assess this question quantitatively.

The basic framework for computing probabilities in a large landscape of vacua is reviewed in Sec.~\ref{sec-review}, and the probability distribution $d\mathcal{P}/d\log m_\nu$ is computed in Sec.~\ref{sec-calc}. In the remainder of this introduction, we will describe the key physical effects that enter into the analysis, and we will present our main results. \\

\noindent{\bf Summary:~~} There are two competing effects that determine the neutrino mass sum. We assume that the statistical distribution of neutrino masses among the vacua of the landscape favors a large neutrino mass sum, with a force of order unity or less (see Sec.~\ref{sec-force} for the definition of the multiverse force). 

If the anthropic approach is successful, we must demonstrate a compensating effect: that neutrino masses much greater than the observed value are not frequently observed. That is, we must multiply the prior probability for some value of $m_\nu$ by the number of observers that will be produced in regions where $m_\nu$ takes this value. Observers are usually represented by some proxy such as galaxies. We consider two models for observations: at any given time, their rate is proportional to the number of Milky Way-like galaxies, or proportional to the growth rate of this galaxy population (see Sec.~\ref{sec-anthropic}). We sum this rate over a spacetime region called the causal patch~\cite{Bou06} (a standard regulator for the divergent spacetime that results from a positive cosmological constant; see Sec.~\ref{sec-measure}). The product of prior distribution and the abundance of galaxies yields a predicted probability distribution. As usual, if the observed value lies some number of standard deviations from the mean of the predicted distribution, we reject the model (in this case the anthropic approach to $m_\nu$) at the corresponding level of confidence.

The neutrino mass spectrum---the individual distribution of masses among the three active neutrinos---has a noticeable effect on structure formation. We consider two extreme cases. In the {\em normal hierarchy}, one neutrino contributes dominantly to the mass sum $m_\nu$; here we approximate the remaining two as massless. In this case the observed mass splittings require $m_\nu \geq 58$\,meV. In the {\em degenerate hierarchy}, each mass is of order $m_\nu/3$ (and here we approximate them as exactly equal). This case will soon be tested by cosmological observations, since the observed mass splittings would imply $m_\nu\gtrsim 150$\, meV, near the present upper limit. We do not explicitly consider the intermediate case of an inverted hierarchy, with two nearly degenerate massive neutrinos and one light or massless neutrino.

The main challenge lies in estimating the galaxy abundance as a function of $m_\nu$. The effects of one or more massive neutrinos on structure formation are somewhat complex; hence, we compute the linear evolution of density perturbations numerically using Boltzmann codes CAMB~\cite{LewCha99} and CLASS~\cite{BlaLes11}, wherever possible. We will now summarize the key physical effects. A more extensive summary and analytic approximations are given in Sec.~\ref{sec-calc} and Appendix~\ref{sec-analytic}; we recommend Refs.~\cite{LesPas06,LesMan13} for detailed study.

\begin{figure}[t]
\centering
\includegraphics[width=3.25in]{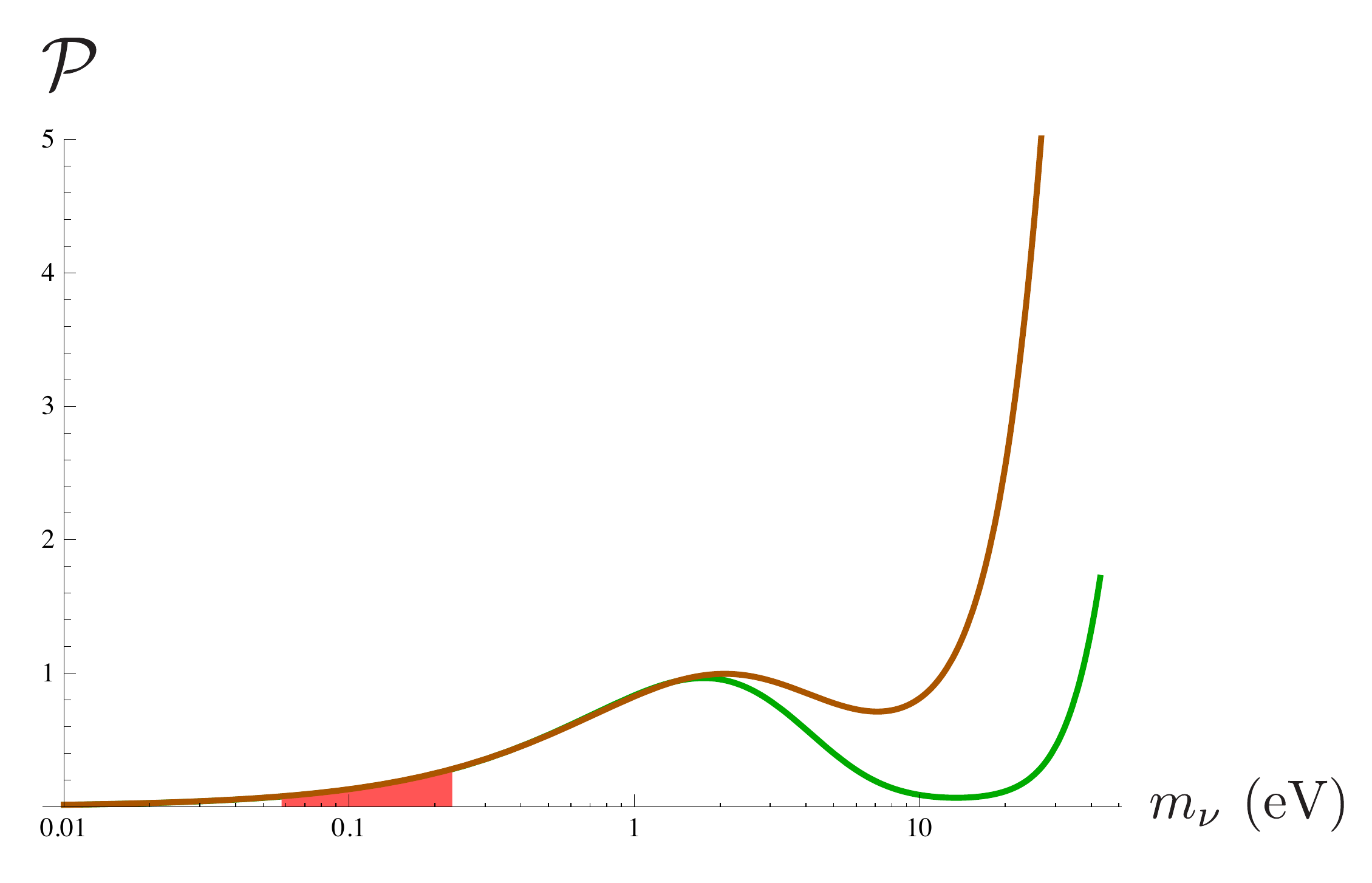}
\caption{No cooling boundary: if one assumed that observers trace dark matter halos of mass $10^{12} M_\odot$ or greater, one would find a bimodal probability distribution over the neutrino mass sum $m_\nu$. This distribution is shown here for a normal hierarchy (orange/upper curve) and degenerate hierarchy (green curve). The range of values $m_\nu$ consistent with observation ($58 \mbox{ meV}<m_\nu<0.23 \mbox{ eV}$, shaded in red) is greatly disfavored, ruling out this model.---By contrast, we shall assume here that observers trace galaxies. Crucially, we shall argue that for $m_\nu \gtrsim 10$\,eV, galaxies do not form even though halos do. This novel catastrophic boundary excludes the mass range above $10$\,eV, leading to a successful anthropic explanation of the neutrino mass (see Fig.~\ref{fig-probdn}).}
\label{fig-ProbNoCutoff}
\end{figure}
After becoming nonrelativistic, neutrinos contribute approximately as pressureless matter to the Friedmann equation. However, they contribute very differently from cold dark matter (CDM) to the growth of perturbations, because neutrinos are light and move fast. This introduces a new physical scale into the problem of structure formation: the  {\em free-streaming scale}\/  is set by the distance over which neutrinos travel until becoming nonrelativistic. It is roughly given by the horizon scale when they become nonrelativistic, with comoving wavenumber $k_{\rm nr}$ (see Appendix~\ref{sec-neutrino} for more details). On this and smaller scales, $k\gtrsim k_{\rm nr}$, neutrinos wipe out their own density perturbations. More importantly, as a nonclustering matter component they change the rate at which CDM perturbations grow, from linear growth in the scale factor ($\delta\propto a$) on large scales, to sub-linear growth on  smaller scales $k\gtrsim k_{\rm nr}$. This suppresses the CDM power spectrum at small scales, see Fig.~\ref{fig-sigmapower}a.

The linear quantity most closely related to the abundance of dark matter halos on the galactic scale $R$ is not the dimensionless CDM power spectrum $k^3 P_{cc}(k)$. Rather, halo abundance is controlled by the smoothed density contrast $\sigma_R$, which is approximately given by the integrated power,
\begin{equation}
\sigma_R\sim \int_0^{1/R} \frac{dk}{k}~ k^3 P_{cc}(k)~,
\label{eq-srintro}
\end{equation}
up to the wavenumber corresponding to the relevant scale. (A more precise formula is given in the main text, where we also describe in detail how halo abundance is computed from $\sigma_R$ using the Press-Schechter formalism.) This distinction turns out to be crucial for large neutrino masses. 

We see from Fig.~\ref{fig-sigmapower}a that for small neutrino masses $m_\nu \lesssim 10$\,eV, the integrand $k^3 P_{cc}(k)$ increases monotonically. Hence the integral for $\sigma_R$ is dominated by its upper limit, i.e., by the power on the scale $k_{\rm gal}\sim 1/R$. This yields the ``bottom-up'' scenario of hierarchical structure formation familiar from our own universe: small halos typically form first, and more massive halos virialize later.

However, for $m_\nu\gtrsim 10$\,eV, the small scale power becomes so suppressed that the dimensionless power spectrum develops a maximum at the free-streaming scale $k_{\rm nr}<k_{\rm gal}$. In this regime, the smoothed density contrast $\sigma_R$ on galactic scales $R$ is no longer dominated by the power at wavenumber $\sim 1/R$. Instead, the power at larger scales than $R$ contributes dominantly to $\sigma_R$. This results in a top-down scenario, where halos first form on cluster scales, nearly simultaneously with galactic-scale halos.

The transition from bottom-up to top-down structure formation around $m_\nu\approx 10$\,eV has not (to our knowledge) been noted in the context of anthropic explanations of the neutrino mass sum. We find here that it is crucial to the analysis, for two reasons. First, it implies that the scales that dominantly contribute to $\sigma_R$ are unaffected by free-streaming for $m_\nu \gtrsim 10$\,eV. Therefore, {\em increasing $m_\nu$ beyond} $\sim 10$\,eV {\em does not suppress CDM structure.} In fact, we find that $\sigma_R$ increases in this range (Fig.~\ref{fig-sigmapower}b). The second implication works in the opposite direction: in the top-down scenario that arises for $m_\nu \gtrsim 10$\,eV, {\em galaxies will not form inside halos at an abundance comparable to our universe}.

\begin{figure*}[t]
\centering
\includegraphics[width=6.5in]{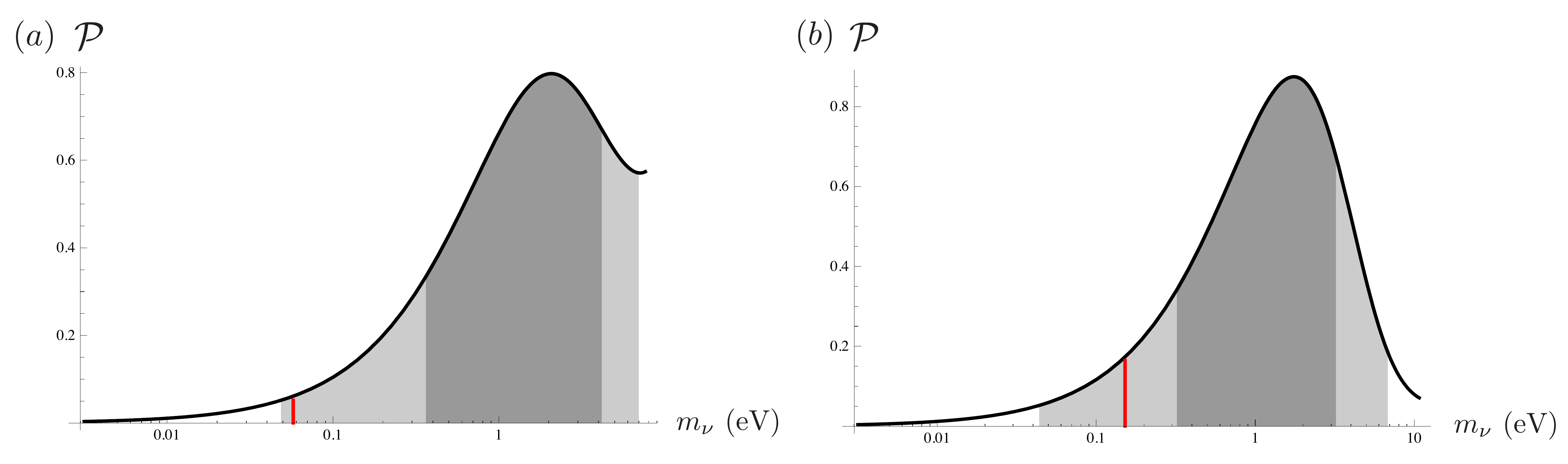}
\caption{Our main result: the probability distribution over the neutrino mass sum $m_\nu$ for (a) a normal hierarchy and (b) a degenerate hierarchy, assuming that observers require galaxies. The plot is the same as Fig.~\ref{fig-ProbNoCutoff}, but the mass range is cut off at $7.7$\,eV ($10.8$\,eV) in the normal (degenerate) case. For greater masses, the first halos form late and are of cluster size; we argue that galaxies do not form efficiently in such halos. We use the second observer model described in Sec.~\ref{sec-anthropic}; results look nearly identical with the first model. We assume a flat prior over $m_\nu$ (see Fig.~\ref{fig-statistics} for other priors).---The central $1\sigma$ and $2\sigma$ regions are shaded. Vertical red lines indicate the lowest possible values for the neutrino mass consistent with available data: $m_{\rm{obs}}\approx 58\mbox{ meV}$  for a normal hierarchy, and $m_{\rm{obs}}\approx 150\mbox{ meV}$ for a degenerate hierarchy. We find that these values are within $2\sigma$ of the median. The agreement would further improve with a less conservative treatment of the detrimental effects of neutrinos on gas cooling in halos, and/or the cosmological detection of a neutrino mass sum larger than the minimal value.}
\label{fig-probdn}
\end{figure*}
Let us discuss each of these implications in turn. We begin by pretending that the stellar mass per halo mass is unaffected by $m_\nu$; in particular, let us suppose that there is no dramatic suppression of star formation in the top-down regime, $m_\nu \gtrsim 10$\,eV. If so, we would be justified in regarding halos as a fair proxy for observers. Here we consider $10^{12} M_\odot$ halos~\cite{MarSha97}. From Fig.~\ref{fig-sigmapower}b, we see that halo abundance decreases with $m_\nu$ up to $m_\nu \sim 10$\,eV; then it begins to increase. Combining this with the assumed prior distribution that favors large $m_\nu$, we would find the probability distribution over $m_\nu$ is bimodal (Fig.~\ref{fig-ProbNoCutoff}). The first peak is at $m_\nu\approx 1$\,eV, followed by a minimum near $10$\,eV and a second peak at much greater mass.\footnote{Fig.~\ref{fig-ProbNoCutoff} does not show the entire peak since CAMB gives results only for $m_\nu \lesssim 40$\,eV. Absent the earlier catastrophic boundary at $10$\,eV that we will assert, a robust effect that would eventually suppress the probability at large neutrino mass is the smallness of the baryon fraction for $m_\nu \gtrsim 100$\,eV. This would suppress the number of baryons (and hence, observers) in the causal patch~\cite{BouHal13}. It would also impose dynamical obstructions to star formation~\cite{TegAgu05}.}  Therefore, {\em if observers traced dark matter halos with $M\gtrsim 10^{12}M_\odot$}, one should conclude that small neutrino masses are greatly disfavored. Such a result would be in significant tension with the current upper bound of $0.23$\,eV, and it would seem to render an anthropic origin of the neutrino mass sum implausible.

However, our fundamental assumption is that observers trace galaxies, not halos. In some cosmologies including our own, galaxies in turn trace halos; if they do, halos are an equally good proxy. But the change of regime from bottom-up to top-down structure formation for $m_\nu \gtrsim 10$\,eV is catastrophic for galaxy formation.

From observation, we know that stars do not form efficiently in bound structures that are much larger than the mass scale of our own galactic halo, $10^{12} M_\odot$. Heuristically, this can be explained by noting that in halos of this size, the cooling timescale for the baryonic gas is greater than the age of the universe~\cite{Bin77,ReeOst77,WhiRee78,BlaVal92,TegAgu05}. In our universe there are galaxies because galactic halos, which produce stars efficiently, formed earlier than these larger halos, which do not. Clusters inherit galaxies that formed in smaller halos, but they do not have significant star formation themselves. 

In a top-down scenario due to large neutrino mass, however, galactic halos would form much later. They would typically be embedded in larger halos that virialize roughly at the same time, with masses characteristic of galaxy groups clusters---but without many galaxies to inherit. The virial temperature and dynamical timescale relevant for baryon cooling will be set by the largest of the nested halos.  (See Sec.~\ref{sec-gala} and Appendix~\ref{sec-cool} for details.) Therefore, cooling will not be efficient: the top-down scenario produces star-poor dark matter clumps, with most baryons remaining in hot gas. 

As a first approximation for this cooling boundary, {\em we cut off the probability distribution at a value} $m_\nu\sim 10$\,eV {\em that corresponds to the onset of the top-down regime}. This overestimates the amount of galaxies just below the cutoff and underestimates it just above. In future work, we plan to include explicit models for successful galaxy cooling beyond the crude top-down vs.\ bottom-up criterion. This should replace the sharp cutoff by a smooth decay of the probability.

We believe that our argument for a cooling catastrophe is robust, because the transition to a top-down scenario is a drastic change of regime. However, the underlying physics is complicated, involving shocks, complicated cooling functions, fragmentation, and feedback from stars, black holes, and supernovae. Suppose therefore that we are wrong. That is, suppose that at $m_\nu \gtrsim 10$\,eV, some unanticipated combination of processes lead to a stellar mass inside the causal patch that is not much less than in our universe. Then one would find that large neutrino masses are unsuppressed (Fig.~\ref{fig-ProbNoCutoff}), and the observed value of $m_\nu$ cannot be explained anthropically. In this sense, the cooling catastrophe we assert can be regarded as a prediction of the anthropic approach to the neutrino mass. To test this prediction, it will be important to investigate galaxy formation for $m_\nu \gtrsim 10$\,eV using simulations that give an adequate treatment of cooling flows and feedback.\\

\begin{figure*}[t]
\centering
\includegraphics[width=6.5in]{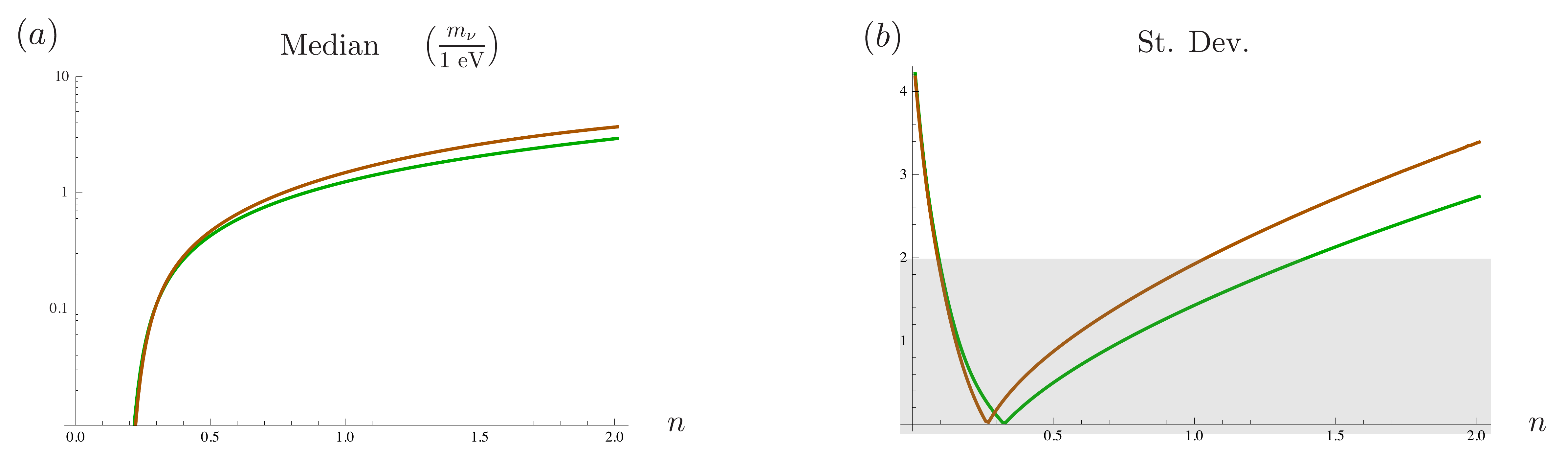}
\caption{The prior distribution of cosmologically produced vacua is assumed to favor large neutrino mass and to have no special feature near the observed magnitude: $d\mathcal{P}_{\rm vac}/d\log m_\nu\propto m_\nu^n$, $n>0$. One then expects $n\sim O(1)$, and the previous two figures all show the case $n=1$. This figure shows that the same conclusions obtain for a considerable range of $n$. (a) The median of the probability distribution as a function of the multiverse force $n$. (b) The standard deviation of the worst case observed value from the median, as a function of $n$. The $2\sigma$ region is shaded. Orange (the upper curve at large $n$ in each plot) is for a normal hierarchy; green is for a degenerate hierarchy.}
\label{fig-statistics}
\end{figure*}

\noindent{\bf Results:~~} Our main results, with the cooling cutoff $m_\nu\lesssim 10$\,eV imposed, are shown in Fig.~\ref{fig-probdn}. We find that the currently allowed range of values for $m_\nu$ is entirely consistent with an anthropic explanation, at better than $2\sigma$. Fig.~\ref{fig-statistics} shows that that our approach succeeds for a wide range of prior distributions $d\mathcal{P}_{\rm vac}/d m_\nu\propto m_\nu^{n-1}$: assuming a normal (degenerate) hierarchy, $m_{\rm obs}$ lies within $2\sigma$ of the median if $0.09<n\leq1.0$ ($0.09<n<1.4$). 

Our chief conclusion is that the neutrino mass sum can be anthropically explained, but only if detrimental effects of neutrinos on galaxy and star formation (rather than halo formation) already become significant at or below $m_\nu\approx 10$\,eV. 

Our results favor larger $m_\nu$ than the minimum values allowed by the observed mass splittings, and in particular they favor a degenerate over a normal hierarchy. Since the observed range is consistent within $2\sigma$ in either case, these are mild preferences rather than sharp predictions. 

There are however two additional reasons why a degenerate hierarchy appears more natural in the context of the anthropic approach. First, with a normal hierarchy one might expect that each neutrino mass scans separately with prior $n_i$. Each prior would have to be assumed positive and $O(1)$. The prior for $m_\nu$ would then be $n=\sum n_i$, and it becomes less plausible that $n$ should be small enough to render the anthropic prediction compatible with observation. A degenerate hierarchy, on the other hand, may be the result of some flavor symmetry that links the masses of the individual neutrinos, leaving only a single scanning parameter. Then it is more plausible that $n$ is small enough to include the observed $m_\nu$.  \looseness=-1

The second reason to prefer a degenerate hierarchy is that it eliminates a viable anthropic window where two neutrinos are extremely massive. If each neutrino has mass of order MeV or greater, neutrons would be stable, leading to a (catastrophic) helium-dominated universe~\cite{TegVil03}. But neutrons will be unstable and the catastrophe is averted, if one neutrino remains light and only the other two become very heavy. With a normal or inverted hierarchy, one has to explain why the one or two heavy neutrinos did not end up in the extremely large mass range above the MeV scale. This can be resolved by assuming that the prior distributions for the individual neutrino masses do have a feature between the eV and the MeV scale, such that the much larger scale is disfavored. With a degenerate hierarchy, this problem does not arise in the first place, since either all neutrinos are light or all are heavy.\\

\noindent{\bf Relation to earlier work:~~} 
Our analysis builds on the pioneering work of Tegmark, Vilenkin and Pogosian~\cite{TegVil03,PogVil04} (see also~\cite{Aguirre:2004qb,Pogosian:2006fx}), who were the first to argue that the neutrino mass admits an anthropic explanation. We agree with their conclusion, but we claim here that the nature of the relevant catastrophic boundary was not correctly identified.

\begin{figure}[t]
\centering
\includegraphics[width=3.25in]{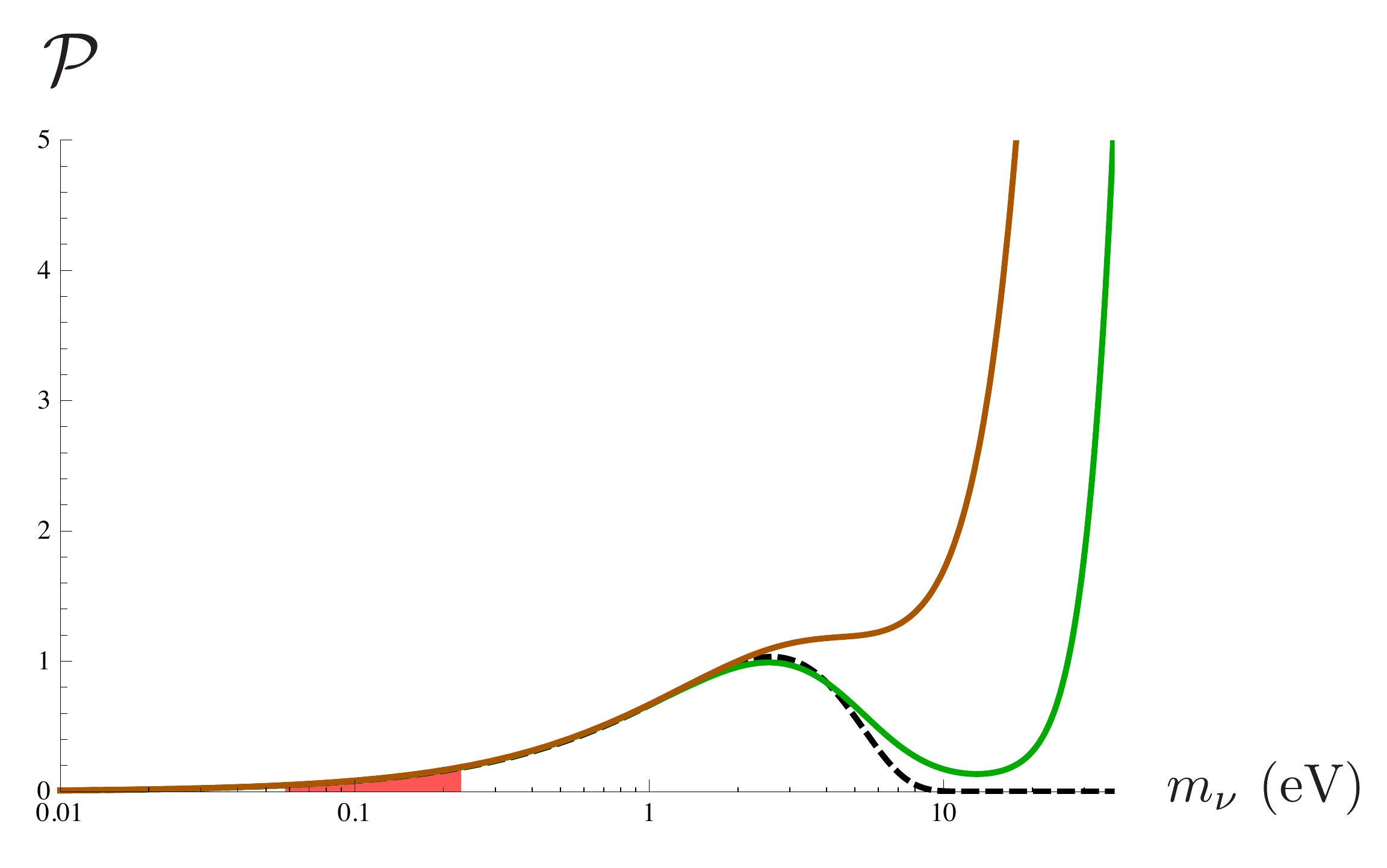}
\caption{The dashed black line shows the probability distribution found by Tegmark {\em et al.}~\cite{TegVil03} for a flat prior over $m_\nu$. This result would seem to remain compatible at about $2\sigma$ with current observational constraints (red shaded region). However, the analytic fitting function for the density contrast $\sigma_R$ used in~\cite{TegVil03,PogVil04} underestimates $\sigma_R$ above a few eV. The solid curves show the probability distribution that results when $\sigma_R$ is computed numerically from Boltzmann codes: orange/upper=normal hierarchy; green/lower=degenerate hierarchy. They differ slightly from Fig.~\ref{fig-ProbNoCutoff} because Ref.~\cite{TegVil03} used a different measure and observer model. Either way, a successful anthropic explanation of $m_\nu$ requires the identification of a catastrophic boundary at or below $10$\,eV.}
\label{fig-ProbVTP}
\end{figure}

Ref.~\cite{TegVil03} does not justify its restriction to the region $m_\nu\lesssim10$\,eV. Moreover, it employs an analytic approximation to $\sigma_R$ that greatly underestimates the halo abundance for $m_\nu\gtrsim 5-10$\,eV. With this approximation, the probability distribution appears to vanish near $10$\,eV due to a paucity of CDM structure; see Fig.~\ref{fig-ProbVTP}. Thus, suppression of CDM structure due to massive neutrinos---rather than the obstruction to cooling at $m\gtrsim 10$\,eV---would appear to provide the relevant catastrophic boundary underlying the anthropic explanation of the neutrino mass sum. \looseness=-1

Here we go further in two respects: our numerical computations show that CDM structure becomes unsuppressed for $m_\nu\gtrsim 10$\,eV. Hence, if neutrino masses have an anthropic origin, a different catastrophic boundary is relevant. And we identify a specific physical effect, the transition to a top-down regime, which had not been noted and which supplies a suitable boundary by suppressing galaxy formation. 

The analytic approximation in question is Eq.~(5) in Ref.~\cite{TegVil03}. It assumes that massive neutrinos suppress the smoothed density contrast $\sigma_R$ on galactic scales by the same factor by which they suppress the matter power on galactic scales. This is accurate for small neutrino masses, because in a bottom-up scenario the shortest scales dominate the integral for $\sigma_R$. The approximation underestimates the abundance of dark matter halos for $m_\nu \gtrsim 10$\,eV, because in this regime $\sigma_R$ is dominated by power at larger scales, which is relatively unsuppressed by free-streaming. More details can be found after Eq.~(\ref{eq-srintro}) and in Sec.~\ref{sec-gala}. 

The discrepancy is revealed by explicit numerical computation of the smoothed density contrast $\sigma_R$ on galactic scales from Boltzmann codes (see Fig.~\ref{fig-sigmapower}). One also finds significantly different results for a normal versus degenerate hierarchy, a distinction that was suppressed in the analytical approximation of Ref.~\cite{TegVil03}.

When the halo abundance is correctly computed, the need for a novel catastrophic boundary at or before $10$\,eV becomes evident (Fig.~\ref{fig-ProbVTP}). Without it, the probability distribution would strongly disfavor small neutrino masses. It would be in significant tension with the an upper bound of $0.23$\,eV or even $1$\,eV, and it would seem to render an anthropic origin of the neutrino mass sum implausible.  

Our computation of $\sigma_R$ from Boltzmann codes, and our compensating identification of a novel catastrophic boundary at $10$\, eV are the main differences to Ref.~\cite{TegVil03}. Another difference is that we use the causal patch measure~\cite{Bou06} to regulate the infinities of eternal inflation. Refs.~\cite{TegVil03,PogVil04} used a different measure that is no longer viable; see Sec.~\ref{sec-measure} for details. This has a visible but comparatively small effect on the probability distribution: by comparing Fig.~\ref{fig-ProbNoCutoff} with Fig.~\ref{fig-ProbVTP}, one sees that the causal patch is somewhat more favorable to an anthropic explanation of $m_\nu$. The causal patch also renders more robust our conjecture that star formation is ineffective for $m_\nu\gtrsim 10$\,eV, as discussed in more detail later. 

We also build on the seminal investigation of catastrophic boundaries in cosmology by Tegmark {\em et al.}~\cite{TegAgu05} (see also Ref.~\cite{BouHal09}), who emphasized the crucial role of cooling. We believe that our present work is the first to associate catastrophic cooling failure to a top-down structure formation scenario. Ref.~\cite{TegVil03} points out a number of distinct catastrophies at very large neutrino mass: For example, neutrinos act as cold dark matter for $m_\nu \gg 100$\,eV, which also may be detrimental to star formation. (However, this does not counteract the abundance of CDM structure we find at $m_\nu \gtrsim 10$\,eV. Fig.~\ref{fig-ProbVTP} illustrates that a cutoff at any scale larger than $10$\,eV, say at $m_\nu \approx 30$\,eV, would make small neutrino masses too improbable for an anthropic explanation to work.)

\section{Predictions in a Large Landscape}
\label{sec-review}

If a theory has a large number of metastable vacua, most predictions will be statistical in nature. We are usually interested in understanding the magnitude of a particular parameter $x$, such as the cosmological constant or in the present case, the neutrino mass; hence we wish to compute a probability distribution $\frac{d\mathcal{P}}{d\log x}$.

Fundamentally, the probability $d\mathcal{P}$ is proportional to the number of observations $dN_{\rm obs}$ that find the parameter to lie in the range $(\log x, \log x+d\log x)$. Thus, our task is to compute $dN_{\rm obs}$. This can be done by weighting a prior probability distribution $ f(x)$, which comes from the underlying theory, by the number $w(x)$ of observations that will be made in a vacuum where $x$ takes on a particular value:
\begin{equation}
\frac{d\mathcal{P}}{d\log x} = w(x) f(x)~.
\label{eq-pwtp}
\end{equation}
We will discuss each factor in turn. Our presentation in each subsection will be general at first, before specializing to the case of the neutrino mass, $x=m_\nu$.

\subsection{Prior as a Multiverse Probability Force}
\label{sec-force}

The prior is defined by
\begin{equation}
 f(x) = \frac{d N_{\rm vac}}{d\log x}~,
\end{equation}
Here $x$ will be a parameter in the effective theory at low energies whose scale $\log x$ one would like to predict or explain; $d N_{\rm vac} =  f(x) d\log x$ is the number of long-lived metastable vacua\footnote{Strictly, what matters is not the abundance of such vacua in the effective potential but in the multiverse: cosmological dynamics could favor the production of some vacua over others. For most low-energy parameters one expects that such selection effects are uncorrelated with $x$ in the range of interest. In any case, we shall take the prior $ f$ to be an effective distribution that incorporates cosmological dynamics.} in which the parameter takes on values in the range $(\log x, \log x+d\log x)$.

With the notable exception of the cosmological constant, the prior distribution for most parameters is not well known. This is a technical problem: in the string landscape, $f(x)$ should in principle be computable. In practice, it is difficult to derive phenomena far below the fundamental scale (the Planck or string scale) directly. However, this need not be an obstruction to progress, any more than the fact that we cannot derive the Standard Model from a more fundamental theory prevented us from discovering it. 

Consider an arbitrary low-energy parameter $x$. In any large landscape the prior distribution $f(x)$ should admit an effective description~\cite{Don97,HalNom07,BouHal09}. To avoid putting in the answer, one may assume that $f(x)$ has no special features (such as a maximum) in a wide logarithmic range of values. This range should include but be much larger than the range compatible with observation. One can then parametrize the prior distribution by a ``statistical pressure'' or ``multiverse probability force'' towards large or smaller values,
\begin{equation}
n \equiv \frac{d\log f}{d\log x}~,
\end{equation}
where $n$ is approximately constant. For example, a flat prior distribution over $\log x$ corresponds to $n=0$. If the prior is flat over $x$, $d N_{\rm vac}/dx=$ const., then $n=1$.

Suppose that there is a regime change sufficiently near the observed value $\log x_0$, such that the number of observers (or at least, of observers like us) $w(x)$ drops dramatically above or below a critical value $\log x_c$. Suppose for definiteness that $x_0\lesssim x_c$. If the probability force favors large values of $x$, but not too strongly [$n>0$, $n\sim O(1)$], then the observed value can be explained. Similarly, with a negative probability force, one can explain the proximity of $x$ to nearby catastrophic boundary at some smaller $x_c\lesssim x_0$. 

Recent successful examples of this approach include an explanation of the coincidence that dark and baryonic densities are comparable~\cite{BouHal13}, the fine-tuning of the weak scale~\cite{AgrBar98,BouHal13,HalPin14}, and the comparability of several large, a priori unrelated timescales in cosmology~\cite{BouHal09}.  In each case, the required assumption about the probability force is weak and qualitative: $n\sim \pm O(1)$. Thus phenomenological models of the landscape have significant explanatory value, while constraining the underlying prior distributions through the sign (and roughly the strength) of the probability force $n$. It is particularly instructive to keep track of the combination of (and possible conflict between) forces $n_i$ needed to simultaneously explain multiple parameters $x_i$ ~\cite{BouHal09,HalPin14,DerHal14}.

Now let us turn to the prior for the total neutrino mass, $\mathcal{P}_{\rm vac}(m_\nu)$. We know of no physical reason why a {\em minimum} neutrino mass should be necessary for observers. Hence, to obtain a normalizable probability distribution $f$, we must assume that the effective prior distribution favors large $m_\nu$:
\begin{equation}
\frac{d \mathcal{P}_{\rm vac}}{d\log m_\nu} \propto m_\nu^n~,~~ n>0~,
\label{eq-npos}
\end{equation}
in some large logarithmic neighborhood of the observed value, $\sim0.1$\,eV. A natural and simple choice is $n=1$, and we will use this value for definiteness in most plots. More generally, we will find that a comfortable range of values $0<n\approx O(1)$ is consistent with an anthropic explanation of the neutrino mass, but not a value much greater than $1$ (see Fig.~\ref{fig-statistics}).

\subsection{Anthropic Weighting}
\label{sec-anthropic}

The probability distribution over $\log x$ relevant for comparing the theory with observation is obtained by conditioning $\tilde p$ on the presence of observers. More quantitatively, one weights by the number of observations 
\begin{equation}
w(x) = dN_{\rm obs}/d N_{\rm vac}
\end{equation}
that are made in a vacuum where $x$ takes on a specific value. Generically, $w(x)$ will be unsuppressed in a large region either above or below the observed value, or both. Thus, the anthropic factor is not doing all the work; the prior distribution is crucial for comparing the theory to observation.

In this paper we will consider two different models for the number of observations $w(x)$ that are performed in the universe. Both are based on the assumption that observers require galaxies, say of halo mass comparable to the Milky Way's, $10^{12}$ solar masses. The first model assumes that the rate at which observations occur in a given spatial region per unit proper time, $\dot w(x)$, is proportional to the total mass $M_{\rm gal}$ of such galaxies, at every instant; hence
\begin{equation}
w(x) = \int dt~M_{\rm gal}(t)~~~~~\text{(Observer~Model~1)}~.
\label{eq-om1}
\end{equation}
The second model (which reduces to the choice made in~\cite{TegVil03}) assumes instead that the rate of observation is proportional to the rate $\dot M$ at which the above total galaxy mass grows:
\begin{equation}
w(x) = \int dt~\dot M_{\rm gal}(t)~~~~~\text{(Observer~Model~2)}~.
\label{eq-om2}
\end{equation}
The two models can be thought of as two different approximations taken to an extreme. In the first, observations would be made continuously in the galaxy, at fixed rate per unit stellar mass, no matter how old the stars become. In the second model, observations would occur instantaneously as baryons cool and form stars; no observations would be assigned to a galaxy that is not growing. (The second model was used in Ref.~\cite{TegVil03}; note that in the context of the measure used there the integral over time is trivial, yielding the collapse fraction $F_R$.) The truth is likely somewhere in between the two models. However, we will find that our results depend only weakly on the model, so we expect our conclusions to be robust.

\subsection{Measure}
\label{sec-measure}

A cosmology with at least one long-lived de~Sitter vacuum gives rise to eternal inflation: the universe will grow without bound and remain at finite temperature in arbitrarily large volumes at late times. Hence, all possible events will occur infinitely many times. This applies in particular to observations. Thus a regulator or ``measure'' must be introduced to obtain a finite anthropic factor $w(x)$. For this problem to exist, it is not necessary that the theory predict a large landscape; one de~Sitter vacuum (such as, apparently, ours~\cite{Rie98,Per98}) is enough. But the measure problem becomes particularly glaring in the landscape context: globally, every type of vacuum bubble is produced infinitely many times, and each bubble universe contains an infinite comoving volume. 

Existing analyses of the anthropic origin of neutrino masses preceded a period of significant progress on the measure problem of eternal inflation. Following Weinberg~\cite{Wei87}, Refs.~\cite{TegVil03, PogVil04} regulate the divergences of the cosmological dynamics by estimating the number of observers {\em per baryon}. This measure can no longer be considered viable~\cite{Pag06,BouFre06b}. Note, however, that our choice of measure is {\em not} responsible for the main differences between our results and those of~\cite{TegVil03}, as described at the end of Sec.~\ref{sec-intro}.

In this paper, we will use the causal patch measure~\cite{Bou06}, which regulates eternal inflation by considering a single causally connected region and averaging over its possible histories. This proposal is very generally defined, requiring only causal structure. It is also well motivated: it merely applies to cosmology an existing restriction that was already needed for the unitary evaporation of black holes~\cite{SusTho93}. Though proposed on formal grounds, the causal patch has met with phenomenological success; two examples are described in Appendix~\ref{sec-patch}. We take this as evidence that it approximates the correct measure well (at least in regions with positive cosmological constant~\cite{Sal09}).

A potential landscape is consistent with the observed cosmological history only if it is multi-dimensional with large energy differences between neighboring vacua~\cite{Bou12}.\footnote{The decay of our parent vacuum must release enough energy to heat our universe at least to the temperature of big bang nucleosynthesis, which requires $\Delta \Lambda\gg 1$\,(MeV)$^4$. This is the reason why a multidimensional landscape is essential. One-dimensional ``washboard'' landscapes~\cite{Abb85,BroTei87,BroTei88} are ruled out, because they must have $\Delta\Lambda\lesssim 10^{-35}$\,(MeV)$^4$ so as to naturally include at least one vacuum like ours.} String theory gives rise to such a structure upon compactification to three spatial dimensions~\cite{BP}, with $\Delta\Lambda$ not much below unity. 

The causal patch will contain a particular decay chain through de~Sitter vacua in the landscape, ending with a big crunch in a vacuum with negative cosmological constant; each such chain is weighted by its probability, i.e., by the product of branching ratios~\cite{Bou06}. For a typical decay chain, none of the vacua will have anomalously small cosmological constant $\Lambda\ll \Delta \Lambda$. Thus, after conditioning on observers, there will be one vacuum with small cosmological constant in the causal patch, and we need only be concerned with how the causal patch regulates the volume of the corresponding bubble universe. 

Here we focus on the variation of the neutrino mass only, so we shall take this vacuum to be otherwise like ours. In particular we set the cosmological constant to the observed value,  $\Lambda\sim 10^{-123}$, and we take the spatial geometry to be flat. The metric is of the Friedman-Robertson-Walker (FRW) type:
\begin{equation}
ds^2 = -dt^2 +a(t)^2 (dr^2 + r^2 d\Omega^2)~,
\end{equation}
where $a$ is the scale factor, $r$ is the comoving radius, $t$ is proper time, and $d\Omega^2$ is the metric on the unit two-sphere.

By definition, the causal patch is the causal past of the future endpoint of a geodesic; thus its boundary consists of the past light-cone of such a point. We are interested in the boundaries of the causal patch during the time when a long-lived de Sitter vacuum still contains matter. A future decay has an exponentially small effect on the location of the patch boundary at much earlier times, so the patch can be computed by treating the vacuum as completely stable. The patch boundary is thus the cosmological event horizon. Its comoving radius at FRW time $t$ is obtained by tracing a light-ray back from future de~Sitter infinity:
\begin{equation} 
r_{\rm patch}(t) = \int_t^\infty \frac{dt'}{a(t')}~.
\label{eq-rpatch}
\end{equation}
The physical volume of the patch is 
\begin{equation}
V_{\rm phys}(t) = \frac{4\pi}{3} a(t)^3 r_{\rm patch}(t)^3~.
\end{equation}

As described in the previous subsection, we estimate the rate of observations per unit time as proportional to the total mass of all galaxies in the physical volume of the patch (for observer model 1), or to the rate of increase of this mass (for observer model 2).  We can write this quantity as
\begin{equation}
M_{\rm gal}(t) = \rho_{bc}(t) V_{\rm phys}(t) F_{R}(t) G_{R}(t) ~.
\end{equation}
The first two factors give the total mass $M_{bc}$ of baryons and cold dark matter in the patch at the time $t$. The collapse fraction $F_{R}$ is the fraction of this mass that is contained in halos of mass greater than $10^{12} M_\odot$, corresponding to a comoving distance scale $R$: $M_{\rm halo} = M_{bc} F_{R}$. The galaxy fraction $G_{R}$ is the fraction of this latter mass that represents baryons in galaxies, $M_{\rm gal} = M_{\rm halo} G_{R}$. 

Combining this with Eqs.~(\ref{eq-om1}), (\ref{eq-npos}), and (\ref{eq-pwtp}), the (unnormalized) probability distribution over the neutrino mass is given by
\begin{equation}
\frac{dp}{d\log m_\nu}  \propto m_\nu^n \int dt\,(r_{\rm patch}a)^3\,\rho_{bc}\,
F_{R}\,G_{R}~.
\label{eq-masterprob}
\end{equation} 
in the first observer model; we replace $F_{R}G_{R}$ by $\frac{d}{dt} (F_{R}G_{R})$ for the second observer model.\footnote{Note that the time derivative should not be taken of the entire integrand, for this model. The loss of mass across the horizon due to the shrinking comoving volume of the patch does not produce ``negative galaxies'' inside the patch. At some cost in readability, we could have made this more explicit by defining the integrand in Eq.~(\ref{eq-om2}) as the causal patch volume times the rate of change of the average physical density contributed by galaxies.} Factors in the integrand may in general depend on both $m_\nu $ and $t$.

\section{Calculation of $d\mathcal{P}/d\log m_\nu$}
\label{sec-calc}

\subsection{Fixed, Variable, and Time-dependent Parameters}
\label{sec-para}

We will consider a one-parameter family\footnote{It would clearly be of interest to compute the probability distribution over several parameters including the neutrino mass; for examples of multivariate probability distributions in the landscape, see e.g.~\cite{PogVil04}. Each additional scanning parameter is an additional opportunity to falsity the model. But already with one parameter scanning, one can falsify a model, in the usual way: by computing a probability distribution from the theory. If one finds that the observed value is several standard deviations from the mean, the model is ruled out at the corresponding level of confidence.} of cosmologies, differing from our universe only in the total mass of active neutrinos. More precisely, we consider two such families, since we treat the cases of normal and degenerate neutrino hierarchy separately. Thus, we hold fixed all fundamental parameters other than $m_\nu$. In particular, we fix the vacuum energy density, $\rho_\Lambda=\Lambda/8\pi G$, and the spatially flat geometry of the universe (imposed, presumably, by a mechanism like inflation that is uncorrelated with $m_\nu$). We also hold fixed $\chi_b\equiv \rho_b/n_\gamma$ and $\chi_c\equiv\rho_c/n_\gamma$, the masses {\em per photon} of baryons and CDM. These quantities remain invariant under changes of $m_\nu$, since we hold fixed the fundamental processes that produced the observed baryon and CDM abundances.

For the actual values of these parameters, we use the \emph{Planck} TT+lowP+lensing+ext best fit cosmological parameters~\cite{Ade15}; see Table~\ref{tab:parameters}. The best fit assumes a neutrino mass of about $0.06$\,eV~\cite{Ade15}, whereas strictly, one should use a best fit marginalized over $m_\nu$ for the purposes of our paper. However, this has virtually no effect on the fixed cosmological parameters such as $\rho_\Lambda$, $\chi_b$, and $\chi_c$, because neutrinos are already constrained to contribute a very small fraction to the total density. For example, the best-fit for the Hubble parameter\footnote{Unless otherwise specified, we quote the Hubble parameter in units $\mbox{km} \mbox{ s}^{-1} \mbox{ Mpc}^{-1}$ throughout.} (\emph{Planck} TT+lowP+lensing+ext~\cite{Ade15}) shifts from $67.9\pm 0.55$ ($m_\nu\approx 0.06$\,eV) to $67.7\pm 0.6$ (marginalized over $m_\nu$). This difference is negligible compared to current error bars and the discrepancies between different cosmological datasets.

When considering entire cosmological histories, as we do, it is best to specify each cosmology in terms of time-independent parameters such as $\Lambda$, $\chi_b$, $\chi_c$, and $m_\nu$. However, we use Boltzmann codes such as CAMB and CLASS to compute power spectra wherever possible (i.e., for $z\geq 0$). These codes expect input parameters that specify the cosmological model in terms of their present values, at redshift $z=0$. It is not clear what one would mean by the ``present'' time in an alternate cosmology, but for the purposes of CAMB and CLASS, $z=0$ is defined to be the time at which the CMB temperature takes the observed value, $T_{\rm CMB}\approx 2.7$\, K. 

\begin{table}[t]
\caption{The cosmological parameters used in our calculation, as well as the resulting mass per photon of baryons and CDM, $\chi_b$ and $\chi_c$. $T_{\rm CMB}$ is a \emph{Planck} TT+lowP+BAO fit, while all others are from \emph{Planck} TT+lowP+lensing+ext best fit  values. We take $k_{\rm pivot}=0.05 \mbox{ Mpc}^{-1}$.}\label{tab:parameters}
\begin{center}
\begin{tabular}{cc}
\hline \hline
Parameter &  Value \hspace{0.5 in}\\
\hline
 $T_{\rm CMB}$ & $2.722 \mbox{ K}$ \indent\\
 \indent $H_0$  &  $67.90$\indent\\
 $\Omega_{m}$ & $0.3065$\indent\\
 $\Omega_\Lambda$ & $0.6935$\indent\\
 $\Omega_{b} h^2$ &  $0.02227$\indent\\
 $\Omega_{c} h^2$ &  $0.1184$\indent\\
 $10^{9} A_s$ &  $2.143$\indent\\
 $n_s$ & $0.9681$\indent \\
 $\chi_b$ & $0.5745 \mbox{ eV}$\indent\\
  $\chi_c$ & $3.054 \mbox{ eV}$\indent\\
 \hline 
\end{tabular}
\end{center}
\end{table}

Thus we must derive the values of various time-dependent quantities at the time when the universe reaches this temperature, as a function of $m_\nu$, with other time-independent parameters fixed as described above. One finds for the Hubble parameter and the density parameters
\begin{eqnarray} 
H(m_\nu;z=0) & = & H_0\left(\frac{\chi_{bc\Lambda\nu}}{\chi_{bc\Lambda\nu_0}}\right)^{1/2}~, \label{Hubbleeqn}\\
\Omega_X(m_\nu;z=0) & = & \frac{\chi_{X}}{\chi_{bc\Lambda\nu}}~,~~ X \in \{b,c,\Lambda,\nu\} ~.
\end{eqnarray} 
Here multiple indices imply summation, for example $\chi_{bc}\equiv\chi_b+\chi_c$. The fixed parameters $\chi_b$ and $\chi_c$ were defined above. The fixed parameter $\chi_\Lambda\equiv \rho_\Lambda/n_\gamma(z=0)$ is defined for notational convenience as the observed vacuum energy per photon at the {\em present} observed CMB temperature. The $m_\nu$-dependent parameter $\chi_\nu(m_\nu)=\frac{3}{11} m_\nu$ is the neutrino mass per photon. $H_0\equiv H(0.06$\,eV$;z=0)$ and $\chi_{\nu_0}=\chi_\nu(0.06$\,eV$)$ are observed values, corresponding to the {\sl Planck} best fit baseline model.

\subsection{Homogeneous Evolution}
\label{sec-homo}

For computing the volume of the causal patch, the factor $(r_{\rm patch}a)^3$ in Eq.~(\ref{eq-masterprob}), we will need to know the scale factor. Unless structure is present, the integrand will be suppressed by the Press-Schechter factor $F_R$; hence it suffices to use an analytic solution valid to excellent approximation in the matter and vacuum eras:
\begin{equation}
a(m_\nu;t) = \left[ \cot\lambda \, \sinh\left(\frac{3\sin\lambda}{2}\,H_0t \right)\right]^{2/3}~,
\label{eq-asol}
\end{equation}
The solution depends on $m_\nu$ through $\Omega_\Lambda(m_\nu;z=0) \equiv \sin^2\lambda$.\\

Since $\chi_{bc}$ does not depend on $m_\nu$ and $\rho_{bc}=n_\gamma \chi_{bc}$, $\rho_{bc}(z=0)$ does not depend on $m_\nu$. Moreover, since the scale factor in Eq.~(\ref{eq-asol}) is normalized so that $a=1$ at $z=0$, we have  $\rho_{bc}(t) =\rho_{bc}(z=0)/a(t)^3$ for all values of $m_\nu$. Thus Eq.~(\ref{eq-masterprob}) simplifies to
\begin{equation}
\frac{d\mathcal{P}}{d\log m_\nu}  \propto  m_\nu^n \int dt\, r_{\rm patch}^3\,F_{R} G_{R}~,
\label{eq-masterprob2}
\end{equation}
where $r_{\rm patch}$ is given by Eq.~(\ref{eq-rpatch}), and we are dropping $m_\nu$-independent normalization factors as usual. 

\begin{figure}[t]
\centering
\includegraphics[width=3.25 in]{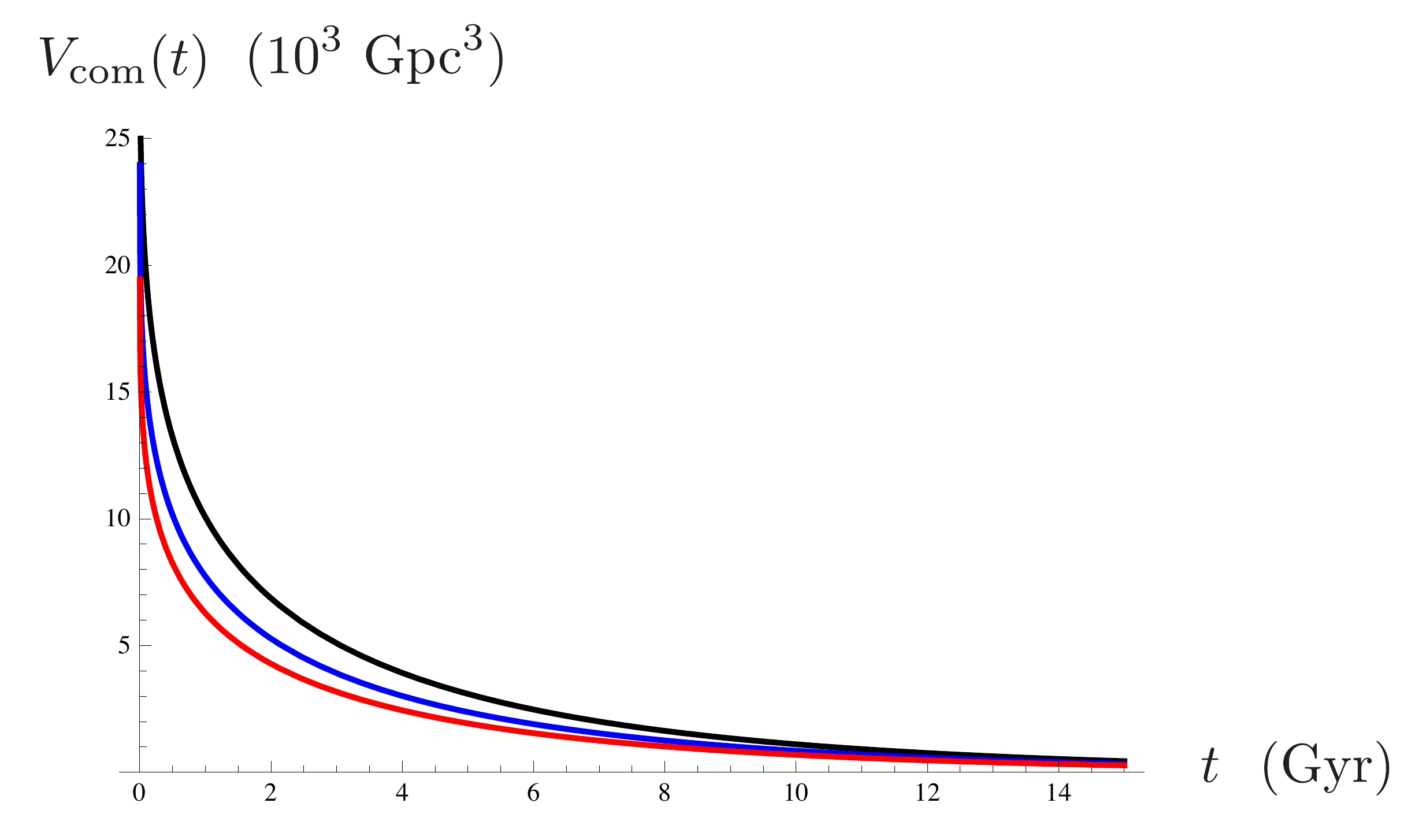}
\caption{The comoving volume of the causal patch for $m_\nu=0$ (black), $m_\nu=4$ (blue), and $m_\nu=8$ (red).}
\label{fig-patchvolume}
\end{figure}

The comoving volume of the causal patch is shown in Fig.~\ref{fig-patchvolume}. We note that already at the homogeneous level, a nonzero neutrino mass is slightly disfavored because it decreases the size of the causal patch. We also note that the patch size is maximal at early times and decreases rapidly. Hence galaxies that form very late effectively fail to contribute to the probability for a given parameter value.

\subsection{Halo Formation}
\label{sec-halo}

The next factor in Eq.~(\ref{eq-masterprob2}) is the collapse fraction $F_{R}(m_\nu,t)$. It captures the effects of neutrinos on structure formation: recall that $F_{R}$ is defined as the fraction of baryonic and cold dark matter that is contained in halos of mass $10^{12} M_\odot$ or greater.  It captures the effects of neutrinos on structure formation. Recall that $F_{R}$ is defined as the fraction of baryonic and cold dark matter in virialized halos of mass scale $10^{12} M_\odot$ or greater. This corresponds to a comoving distance scale $R\sim 1.8 \mbox{ Mpc}$.\footnote{The comoving scale $R$ is independent of $m_\nu$ because $\rho_{bc}(z=0)$ is. However, when expressed in units of Mpc$/h$ it depends on $m_\nu$ through Eq.~(\ref{Hubbleeqn}).} 

The collapse fraction can be determined using the Press-Schechter formalism~\cite{PreSch73}. Before nonlinearities are important, the density contrast\footnote{We use the CDM density contrast and power spectrum to compute the Press-Schechter factor $F$. This matches $N$-body simulations better than using the full matter density contrast including neutrinos~\cite{CasSef13}. It is also a conservative choice, since the total matter power spectrum is further suppressed at large $m_\nu$, by a factor $(1-f_\nu)^2$ below the free streaming scale.} $\delta(x,t)$ smoothed on a scale $R$ has a Gaussian distribution,
\begin{equation}
\mathcal{P}(\delta, t) \, d\delta \sim \exp\left(-\frac{\delta^2}{2\sigma_R^2}\right) d\delta~,
\end{equation}
with standard deviation $\sigma_R(t)$. Fluctuations that exceed a certain threshold $\delta_*\sim O(1)$ in the linear analysis will have become gravitationally bound. Hence,
\begin{equation}  F_{R}(t) = \int_{\delta_*}^\infty \mathcal{P}(\delta, t) d\delta = \mbox{erfc}\left(\frac{\delta_*}{\sqrt{2}\sigma_R(t)}\right)~.\label{anth2}
\end{equation}
We use the canonical value $\delta_*=1.69$, which is obtained by comparing the linear perturbation to a spherical collapse model.\footnote{For structure that forms in the vacuum era, the collapse threshold is slightly lowered~\cite{TegVil03}, whereas in the presence of an appreciable neutrino fraction $\delta_*$ should be slightly increased~\cite{LoV14}. If we adapted $\delta_*$ accordingly, the net effect would be to further suppress structure at large $ m_\nu$, in favor of an anthropic origin of the neutrino mass. However, appropriate values of $\delta_*$ have so far been estimated only for rather small neutrino masses. Ultimately, it would be preferable to sidestep the Press-Schechter approximation altogether. Our analysis could be dramatically improved by using proper $N$-body simulations to compute structure formation, including an adequate treatment of baryonic physics.}

The standard deviation of the smoothed density contrast is given by~\cite{Dod03}
\begin{equation}
\sigma_R^2 \equiv \langle \delta^2_R(x) \rangle~,
\end{equation}
with
\begin{equation}
\delta_R\equiv \int d^3 {\mathbf x'}\, \delta( {\mathbf x} )\, W_R(|{\mathbf x} -{\mathbf x'}|)~,
\end{equation}
where $\delta({\mathbf x})=\delta\rho_c/\rho_c$ is the fractional overdensity of cold dark matter.
We use the {\em top hat} window function, $W_R(x) = 1$ for $|x|\leq R$ and $W_R(x)=0$ otherwise. 

Equivalently, the smoothed density contrast can be computed from Fourier-transformed quantities:
\begin{equation}
\sigma_R^2 = \int_0^\infty  \frac{dk}{k}\frac{k^3 P_{cc}(k)}{2\pi^2}\, |W_R(k)|^2~,
\label{eq-sp}
\end{equation}
where $W_R(k) = \frac{3}{(kR)^3}(\sin{kR} -kR\cos{kR})$. The CDM power spectrum is defined by
\begin{equation}
\langle \tilde\delta({\mathbf k}) \tilde\delta({\mathbf k'})\rangle =
(2\pi)^3 P_{cc}(k) \delta^3({\mathbf k} -{\mathbf k'})~, 
\label{eq-pdelta}
\end{equation}
where $\tilde\delta({\mathbf k})$ is the Fourier transform of $\delta({\mathbf x})$ and $\delta^3$ is the Dirac delta function.

To evaluate $\sigma_R(m_\nu, t)$, we use the CAMB code~\cite{LewCha99} to compute the CDM power spectrum $P_{cc}(k)$ as a function of time, in models with different neutrino mass. We evaluate the integral in Eq.~(\ref{eq-sp}) numerically. We have also checked our results using the CLASS code~\cite{BlaLes11}. We noticed a small discrepancy in the output of $k^3 P_{\rm cc}$ at the largest neutrino masses we consider, $m_\nu \sim 10$\, eV, where CLASS gives a slightly larger amplitude for the free-streaming peak. By lowering the cutoff on $m_\nu$ described in Sec.~\ref{sec-gala}, the CLASS output would only strengthen the anthropic explanation of the observed neutrino mass range.

Available Boltzmann codes do not return power spectra for negative redshifts, that is, for times when the CMB temperature is below $2.7$\,K. {\em In this regime only}, we estimate $\sigma_R$ by extrapolating our numerical results to negative redshifts semi-analytically as described in Appendix~\ref{sec-extrapolate}. This regime is not a dominant contributor to the overall probability distribution, due to the smallness of the causal patch at late times, and since vacuum domination terminates structure formation in any case.

\begin{figure}[t]
\centering
\includegraphics[width=3.25 in]{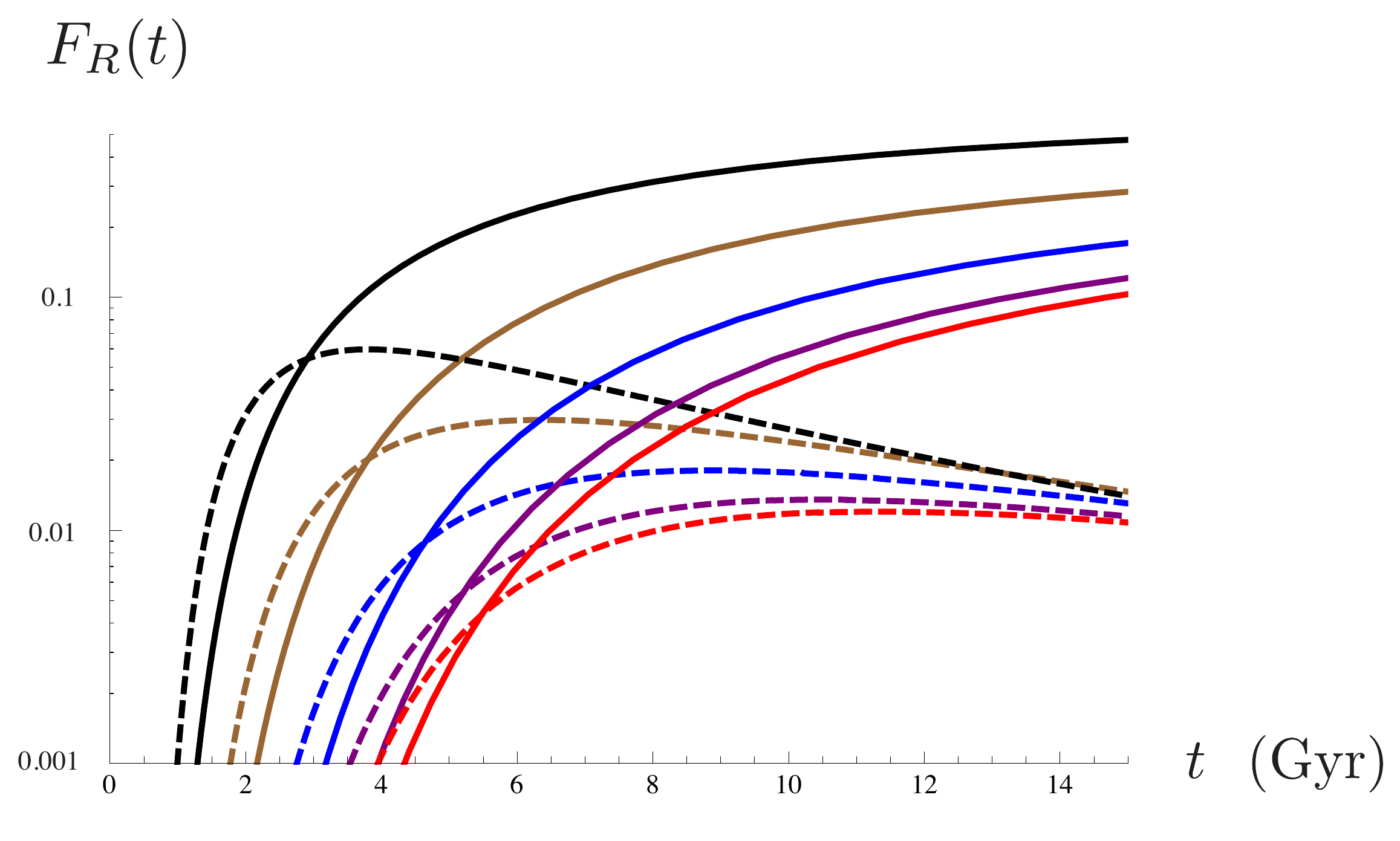}
\caption{The Press-Schechter factor (solid lines) and its derivative (dashed lines) at the galaxy scale, $10^{12} M_\odot$, for a normal hierarchy with $m_\nu=0$\,eV (black), $m_\nu=2$\,eV (brown), $m_\nu=4$\,eV (blue), $m_\nu=6$\,eV (purple) and $m_\nu=8$\,eV (red). Each is used to define either of the two observer models of Sec.~\ref{sec-anthropic}. Note that massive neutrinos suppress structure at all times, but much more so at early times~\cite{Klypin:1992sf,Klypin:1994iu,Mo:1994bc,Kauffmann:1994ty,Ma:1994ub,Liddle:1995ay}.}
\label{fig-pressschechter}
\end{figure}

We compute $F_{R}$ and $\dot F_{R}$ from Eq.~(\ref{anth2}); the results are shown in Fig.~\ref{fig-pressschechter}.

\subsection{Galaxy Formation: Neutrino-Induced Cooling Catastrophe}
\label{sec-gala}

The final factor $G_R(m_\nu,t)$ in Eq.~(\ref{eq-masterprob2}) is the fraction of the halo mass in baryons within galaxies. To approximate this, we must first investigate the effect of a top-down structure scenario (present at $m_\nu\gtrsim 8-10$\,eV, as discussed in Sec.~\ref{sec-intro}) on galaxy formation.

In our universe galaxies form in halos with masses between $10^7 M_\odot$ and $10^{12} M_\odot$. Larger halos can inherit galaxies from mergers, resulting in galaxy groups and clusters, with masses ranging from $10^{13} M_\odot$ to $10^{15} M_\odot$. However, halos in the latter mass range do not themselves produce a significant amount of stars, relative to their total mass. 

This fact can be understood as a consequence of the ability, or failure, of baryons to cool rapidly inside newly formed dark matter halos. (For more detail, see Appendix~\ref{sec-cool} and references given there.) Baryons are shock-heated to a virial temperature $T_{\rm vir}$ when they fall into a large dark matter halo. In order to condense into a galaxy at the center of the halo, the baryons must first shed their thermal energy. Cooling can occur by bremsstrahlung at temperatures large enough to ionize hydrogen, or by atomic and molecular line cooling at the lower temperatures attained in smaller halos. 

Analytically, one can estimate the time it takes baryons to cool, $t_{\rm cool}$. The cooling time grows with the mass of the halo (for large masses), and with the time of its formation. It is also easy to compute the gravitational timescale of the halo, $t_{\rm grav}$, which is somewhat shorter than the time of its formation. 

A good match to observation is obtained by the following criterion. If $t_{\rm cool} < t_{\rm grav}$, then cooling is efficient. A significant fraction of baryons (up to 10\%) is converted into stars. This process occurs rapidly, on a timescale that can be treated as instantaneous compared to the age of the universe when the halo formed.

On the other hand, if $t_{\rm cool} > t_{\rm grav}$, then star formation is limited by the cooling time. In this regime, one would still expect a certain amount of rapid star formation at the dense core of the halo, but this is not seen in observations. (This is known as the cooling flow problem.) Observations do not constrain the possibility that a significant portion of baryons will form stars in the distant future, on a timescale much greater than the age of the universe. This time would greatly exceed $t_\Lambda$. Since the causal patch is of a fixed physical size of order the de~Sitter horizon scale, there will be exponentially few halos left in it at late times. Thus, star formation at very late times does not contribute to the probability of a particular universe. (This sensitivity to the matter content inside the cosmological horizon is a key feature distinguishing the causal patch from other interesting measures, such as the fat geodesic or scale factor time cutoff~\cite{BouFre08b}, and it is responsible for several of the chief successes of the causal patch, e.g.~\cite{BouHar07,Fre08,BouLei09,BouHal09,BouHar10,BouFre10d,BouFre11a,BouHal13}.)  Thus, we may take $t_{\rm cool} < t_{\rm grav}$ as a robust condition for galaxy formation to occur in a newly formed halo.

The cooling function that determines the rate of heat dissipation has a complicated form in the relevant halo mass range (see~\cite{AguTeg04} and references therein). Appendix~\ref{sec-cool} describes two different approximations to $t_{\rm cool}$ and $t_{\rm grav}$ that capture different cooling regimes that halos in our analysis might explore. One finds in either regime that at late times, cooling is inefficient for halo masses above the scale of the Milky Way halo:
\begin{equation}
M_{\rm vir} > 10^{12} M_\odot\,,~ t_{\rm vir}\gtrsim O(\mbox{Gyr})~\implies~\mbox{No Galaxy}
\label{eq-coolcond}
\end{equation}
Importantly, the boundary is consistent with the observation that in our universe, there are no galaxies much larger than the Milky Way.

It would be interesting to implement a more precise version of the above boundary as a cutoff on the {\em time}\/ until galaxy formation is efficient, at any value of $m_\nu$. Massive neutrinos {\em delay} structure formation more dramatically than they suppress it (Fig.~\ref{fig-pressschechter}), so such a cutoff would exclude an appreciable fraction of halos from contributing to galaxy formation even at rather small $m_\nu$. Thus it would lead to a greater suppression of intermediate neutrino masses between $1$ and $10$\,eV, and thus would favor the anthropic approach. Instead, we will argue more conservatively for a cooling cutoff on $m_\nu$ around $10$\,eV. We will now identify a change of regime for $m_\nu \gtrsim 10$\,eV. As we shall see, this transition places the dominant halo population so far into the regime of inefficient cooling, that the above rough estimate suffices to conclude that galaxy formation is highly suppressed.

For $m_\nu \lesssim 8-10$\,eV,  recall that the dimensionless matter power spectrum $k^3 P_{cc}(k)$ increases monotonically with $k$ (see Fig.~\ref{fig-sigmapower}a), and the integral for the smoothed density contrast $\sigma_R$ in Eq.~(\ref{eq-sp}) is dominated by the power at the small galactic scale $R$. In this range, the power spectrum preserves the standard hierarchical structure formation we see in our universe, where low mass halos generally form earlier than more massive ones.  Thus, it is not likely for a $10^{12} M_\odot$ halo to be nested inside a more massive overdensity that collapses at the same time.

Above $m_\nu\approx 8-10$\,eV, neutrinos suppress small scale power so much that the dimensionless power spectrum $k^3 P_{cc}(k)$ develops a maximum near the scale associated with free streaming $k_{\rm nr}$ (Fig.~\ref{fig-sigmapower}a). This corresponds to a mass of order $5-100$ times the scale of the Milky Way halo, roughly the scale of galaxy clusters.\footnote{The peak (the free streaming scale) moves to smaller scales as $m_\nu$ is increased. Eventually it crosses the galaxy scale: for $m_\nu \gtrsim 100$\,eV neutrinos act as cold dark matter. But this does not yield an anthropically allowed region, because the dark matter to baryon density ratio $\zeta$ will be too large. This may be detrimental to disk fragmentation~\cite{TegVil03,TegAgu05}. If the causal patch is used, $\zeta\gg 1$ is robustly suppressed independently of any effects on galaxy and star formation, because the total mass of baryons (and thus of observers) in the patch scales like $(1+\zeta)^{-1}$~\cite{BouHal13}. } It implies that the smoothed density contrast on small scales such as $10^{12} M_\odot$ is no longer dominated by the power at the corresponding wavenumber $k$. Instead, the integral in Eq.~(\ref{eq-sp}) is dominated by the maximum of the integrand, near $k_{\rm nr}$. 

This implies that $10^{12} M_\odot$ overdensities become gravitationally bound at the same time as overdensities on larger scales: a top-down scenario. The virial temperature and cooling time will be set by the largest scale that the $10^{12} M_\odot$ overdensity is embedded in, $M_{\rm vir}\gg 10^{12} M_\odot$. Moreover, for such large halos virialization will occur quite late (see Fig.~\ref{fig-pressschechter}), $t_{\rm vir} \gg 5.3$\, Gyr. Hence, for $m_\nu \gtrsim 8-10$\,eV, the cooling condition in Eq.~(\ref{eq-coolcond}) becomes violated, by a substantial margin. 

Note that this conclusion is insensitive to the halo mass scale we associate with observers. Whether we require $10^{10} M_\odot$ or $10^{12} M_\odot$ halos: if the power spectrum peaks at larger scales, the putative galactic halos will be embedded in and virialize together with perturbations on a mass scale well above $10^{12} M_\odot$, leading to a cooling problem.

Let us summarize these considerations and formulate our cooling cutoff on the neutrino mass. If there exists some large scale $k_*<k_{\rm gal}$ such that $k_*^3 P_{cc}(k_*)>k_{\rm gal}^3 P_{\rm cc}(k_{\rm gal})$, we interpret this as indicating top-down structure formation. Let $m_\nu^{\rm max}$ be the greatest neutrino mass sum for which this criterion is {\em not} met, i.e., the largest neutrino mass compatible with bottom-up structure formation. From Boltzmann codes we find $m_\nu^{\rm max}=7.7$\, eV for the normal hierarchy and  $m_\nu^{\rm max}=10.8$\,eV for the degenerate hierarchy. We have argued that cooling fails substantially in the top-down regime, because the first virialized halos are large and form late. Hence, we treat $m_\nu^{\rm max}$ as a sharp catastrophic boundary. We approximate $G_R$ as a step function that vanishes past this critical mass:

\begin{align} G_R(m_\nu,t)= 
\left\{\begin{array}{c}
1~, \indent m_\nu < m_\nu^{\rm max}\\
\ \ 0~, \indent m_\nu \geq m_\nu^{\rm max}~.
\end{array}\right.
\end{align}
We evaluate the integral in Eq.~(\ref{eq-masterprob2}) numerically using Mathematica. The integration is started before structure begins to form, at redshift $z=12$, when $F_{R}$ is negligible. The integration is terminated deep in the vacuum era when $r_{\rm patch}$ becomes exponentially small. Our final result is described in Sec.~\ref{sec-intro}; see Figures~\ref{fig-probdn} and \ref{fig-statistics}.

\acknowledgments
We thank Niayesh Afshordi, Cora Dvorkin, Lawrence Hall, Wayne Hu, Julien Lesgourgues, Marilena Loverde, David J. E. Marsh, Hitoshi Murayama, Levon Pogosian, Joel Primack, Martin Rees, Uros Seljak, David Spergel, Alexander Vilenkin, and Martin White for very helpful discussions and correspondence. We are particularly grateful to Oliver Zahn for explaining how to use and adapt the CAMB software. This work was supported by the Berkeley Center for Theoretical Physics, by the National Science Foundation (award numbers 1214644 and 1316783), by fqxi grant RFP3-1323, and by the US Department of Energy under Contract DE-AC02-05CH11231. Research at the Perimeter Institute is supported by the Government of Canada through Industry Canada and by the Province of Ontario through the Ministry of Research $\&$ Innovation. The work of CZ is supported by an NSF Graduate Fellowship. 

\appendix

\section{Cosmological Constant and the Causal Patch}
\label{sec-patch}

The cosmological constant offers a nice example of the predictive power of a large landscape, and it also illustrates the advantages of the causal patch measure over competing proposals. In this appendix we review Weinberg's 1987 prediction of a positive cosmological constant~\cite{Wei87}, which has since been confirmed by observation~\cite{Rie98,Per98}. We then turn to the more recent success of the causal patch measure in improving the quantitative agreement with the observed magnitude of $\Lambda>0$ (particularly in settings where the primordial density contrast is also allowed to vary), while eliminating specific anthropic assumptions. The goal is to make contact between an example many readers will be familiar with, and the more general formalism for making predictions in the landscape described in Sec.~\ref{sec-review}.

\subsection{Weinberg's Prediction: $\Lambda\sim t_{\rm vir}^{-2}$}
\label{sec-weinberg}

Because $\Lambda=0$ is not a special value from the point of view of particle physics, the prior distribution over the cosmological constant $\Lambda$ should have no sharp feature near $\Lambda=0$; hence to leading order in a Taylor expansion, $d N_{\rm vac}/d\Lambda\approx$ const. for $|\Lambda|\ll 1$.\footnote{In this Appendix we work in Planck units, $G=\hbar=1$.} Hence we have 
\begin{equation}
 \mathcal{P}_{\rm vac}(\Lambda) \propto \Lambda = \exp(\log \Lambda)~:
\end{equation}
the prior favors large magnitude of the cosmological constant. So far, this is just a restatement of the cosmological constant problem in a landscape setting: among many (nonsupersymmetric) vacua, most will tend to have large $\Lambda$, since precise cancellations between the positive and negative contributions to $\Lambda$ are unlikely.

For $\Lambda>0$, structure formation would be severely diminished if $\Lambda$ was large enough to dominate over the matter density of the universe before the time $t_{\rm gal}$ when density perturbations on the scale of galactic haloes would otherwise become nonlinear. (For negative $\Lambda$ of sufficient magnitude, the universe recollapses too soon.) Crudely, the weighting factor $w(x)$ may be approximated as vanishing for $\Lambda> \rho_{\rm NL}$ and constant for $\Lambda< \rho_{\rm NL}$, where $\rho_{\rm vir}\sim t_{\rm gal}^{-2}$ is the energy density at that time~\cite{Wei87}. A refinement~\cite{MarSha97} models $w(x)$ as the fraction of baryons that enter structure of a specified minimum mass. 

Thus, the resulting distribution $\mathcal{P}(\log\Lambda) =  w f$ peaks around $x\sim -2 \log t_{\rm gal}$. $\mathcal{P}$ is suppressed at larger values of $x$ due to the anthropic factor $w$, and at smaller values of $x$ because the prior probability $f$ is low. The model, proposed by Weinberg in 1987, thus predicted a nonzero cosmological constant not much smaller than $\rho_{\rm NL}$. Just such a value has since been discovered~\cite{Rie98,Per98}. The model could have been ruled out at any level of confidence if, instead of a detection, the observational upper bound on $\Lambda$ had continued to improve, moving ever deeper into the region suppressed by the prior. 

Weinberg's argument had a few shortcomings, which we list here. First, the approach actually favors a somewhat larger value of $\Lambda$; the observed value is small at $2-3\sigma$ depending on the assumptions made about the size of galaxies required by observers. More concerningly, the approach would not appear to be robust against variations of the initial density contrast $Q$. It strongly favors  vacua in which both $Q$ and $\Lambda$ are larger than the observed values, unless the prior for $Q$ favors a small magnitude, or unless there is a catastrophic boundary very close to the observed values of $Q$. Neither of these arguments are easy to make.

\subsection{Causal Patch Prediction: $\Lambda\sim t_{\rm obs}^{-2}$}

In much of the older literature, the divergences of eternal inflation were regulated by computing the number of observers {\em per baryon}. (See the beginning of Sec.~\ref{sec-measure} for a brief discussion of the measure problem, and Ref.~\cite{Fre11} for a review.) This was a reasonable first guess, particularly in the context of a landscape where only the cosmological constant varies. However, it is no longer viable in light of more recent insights~\cite{Pag06,BouFre06b}.

The ratio is not well-defined in a landscape where some vacua may not contain any baryons. Worse, it does not actually regulate all infinities, since a long-lived metastable vacuum with positive cosmological constant (such as ours) will have infinite four-volume in any comoving volume; hence, an infinite number of observers ``per baryon'' will be produced by thermal fluctuations at late times. The number of measures that are well-defined and not clearly ruled out is surprisingly small, and the causal patch measure has had the greatest quantitative success so far (at least~\cite{Sal09} when we are interested in relative probabilities for events in vacua with positive cosmological constant, as we are here). Here we give two examples.

First let us recompute the probability distribution over the cosmological constant, $d\mathcal{P}/d\log\Lambda$ with $\Lambda>0$ using the causal patch. We consider a class of observers that live at the (arbitrary but fixed) time $t_{\rm obs}$; for comparing with out observations, we will choose $t_{\rm obs} = 13.8$\,Gyr. But the causal patch at late times coincides with the interior of the cosmological horizon. Because of the exponential expansion, the average density decreases like $e^{-3t/t_\Lambda}$. If $t_{\rm obs}\gg t_\Lambda\sim \Lambda^{-1/2}\sim O(10)$\, Gyr, no observers will be present in the patch, {\em no matter whether or not galaxies form}. This is a much more stringent cutoff than the suppression of galaxy formation which only sets in for a larger value of $\Lambda$, such that $t_{\rm gal} \gg t_\Lambda$. It agrees very well with the observed value of $\Lambda$, resolving the mild ($2-3\sigma$) tension with Weinberg's estimate. It is unaffected by any increase in the primordial density contrast, since $t_{\rm obs}$ contains Gyr time scales that are not shortened by hastening structure formation. It solves the ``Why Now'' problem directly. And it does all this without making any specific assumptions about the nature of observers, except that they are made of stuff that redshifts faster than vacuum energy. (However, in the present paper we do assume that observers require galaxies.) 

The causal patch can also explain why dark and baryonic matter have comparable abundances: the ``Why Comparable'' coincidence. One makes the qualitative assumption that the dark-to-baryonic density ratio $\zeta$ favors large values. But when $\zeta\gg 1$, the causal patch suppresses baryonic observers by a factor $1/(1-\zeta)$, which counteracts the prior distribution, leading to the prediction that $\zeta\sim O(1)$~\cite{BouHal13}.

\section{Structure Formation with Neutrinos}
\label{sec-analytic}

Our calculation was done almost entirely using Boltzmann codes, not analytic approximations. However, for completeness we summarize here the physical origin of the effects of neutrinos on structure formation. In the final subsection~\ref{sec-extrapolate}, we explain the semi-analytic extrapolation formula we have used to extend the code output to negative redshifts. For excellent in-depth treatments of neutrino cosmology, see Refs.~\cite{LesPas06,LesMan13}.

\subsection{Neutrino Cosmology}\label{sec-neutrino}
Around a second after the big bang at the time of decoupling, neutrinos are frozen out with a Fermi-Dirac distribution whose temperature is set by the primordial plasma. Due to $e^{\pm}$ annihilations that heat up the plasma soon after neutrino decoupling, this temperature differs from the temperature of the CMB, which decouples from the plasma much later: $T_{\nu,0} = (4/11)^{1/3}\, T_{\rm CMB}=1.95$\,K.

The energy density and pressure of a single neutrino with mass $m$ at a fixed time since decoupling is thus approximately given by
\begin{align}
\rho_{\nu} &= 2\int \frac{d^3p}{(2\pi)^3}\frac{\sqrt{p^2+m^2}}{e^{p/T_\nu(z)}+1}~,\\
P_{\nu} &= 2\int \frac{d^3p}{(2\pi)^3}\frac{p^2}{3\sqrt{p^2+m^2}}\frac{1}{e^{p/T_\nu(z)}+1}~,\label{pressure}
\end{align}
where $T_\nu(z)=T_{\nu,0}(1+z)$ is the neutrino temperature as it redshifts from the value set at decoupling.

At early times, neutrinos contribute as radiation and add to the total radiation density as
\be \rho_R=\left[1+\frac{7}{8}\left(\frac{4}{11}\right)^{4/3}N_{\rm{eff}}\right]\rho_\gamma~,\ee
where
\be \rho_\gamma=\frac{\pi^2}{15} T_{\rm CMB}^4~,\ee
and where $N_{\rm{eff}}=3.046$ is the effective number of neutrino species, with a slight deviation from $3$ due to non-thermal spectral distortions from the $e^{\pm}$ annihilations. 

Similarly, the number density of neutrinos per species is set by the CMB number density: 
\be n_\nu = \frac{3}{11} n_{\gamma}~, \ee
where
\be n_{\gamma} = \frac{2\zeta(3)}{\pi^2} T_{\rm CMB}^3~.\ee

Neutrinos become approximately non-relativistic once their thermal energy drops below the relativistic kinetic energy, $3 T_\nu(z) < m_\nu$, which occurs at a redshift $z_{\rm{nr}}$ of\footnote{The non-relativistic transition is far from sudden. The neutrino pressure Eq.~\eqref{pressure} has a non-negligible tail long after the redshift Eq.~\eqref{zNR}, which smears out the transition. We thank J.~Lesgourgues for explaining this point to us.}
\be 1+z_{\rm{nr}}=1991 \left(\frac{m_\nu}{1 \, \rm{eV}}\right)~. \label{zNR}\ee

Well after this transition, the density of non-relativistic neutrinos asymptotes to
\be \rho_\nu = m_\nu n_\nu~,\ee
where $m_\nu$ is the sum of masses of all non-relativistic neutrino species. In terms of this, the neutrino density parameter counting only massive neutrinos is
\be \Omega_\nu = \frac{\rho_\nu}{\rho_\ast}~,\ee
where $\rho_\ast$ is the critical density defined by $H^2=8\pi G \rho_\ast/3$, which gives
\be \Omega_\nu h^2 = \left(\frac{m_\nu}{94.5 \mbox{ eV}}\right)~.\ee

The neutrino free streaming scale is set by the typical distance neutrinos travel thermally up to a given time. Roughly, it is given by the horizon scale at early times and stops growing soon after the neutrinos become nonrelativistic; hence it can be crudely approximated by the horizon scale at the nonrelativistic transition, $k_{\rm nr}$. 

On small scales, there are two effects by which neutrinos suppress structure. The most obvious is that density perturbations will be washed out. Thus, free streaming eliminates the contribution of neutrinos to structure, and thus suppresses the total matter power by a factor $\sim (1-f_\nu)^2$, where \be f_\nu = \frac{\Omega_\nu}{\Omega_m}~.\ee
defines the massive neutrino fraction. Conversely, on larger scales neutrinos will remain confined to the over-dense regions and will behave like cold dark matter.

A secondary but more important effect is that the density of massive neutrinos contribute via the Friedmann equation to the Hubble parameter, which controls the friction term in the growth of matter perturbations. But on short scales, they do not contribute to the source term (the density contrast). Therefore, CDM perturbations grow more slowly in the presence of a nonclustering matter component on short scales~\cite{BonEfs80}:
\begin{align}
&\delta_{\rm{c}} \propto a~, \indent \indent k\lesssim k_{\rm{nr}}~,\nonumber \\
&\delta_{\rm{c}} \propto a^{p}~, \ \, \indent k> k_{\rm{nr}}~,
\label{eq-deltap}
\end{align}
where 
\be p = \frac{-1+\sqrt{1+24(1-f_\nu)}}{4}\approx 1-\frac{3}{5}f_\nu < 1~,\ee
with the last approximation valid in the limit of small neutrino masses.
\begin{figure}[t]
\centering
\includegraphics[width=3.5in]{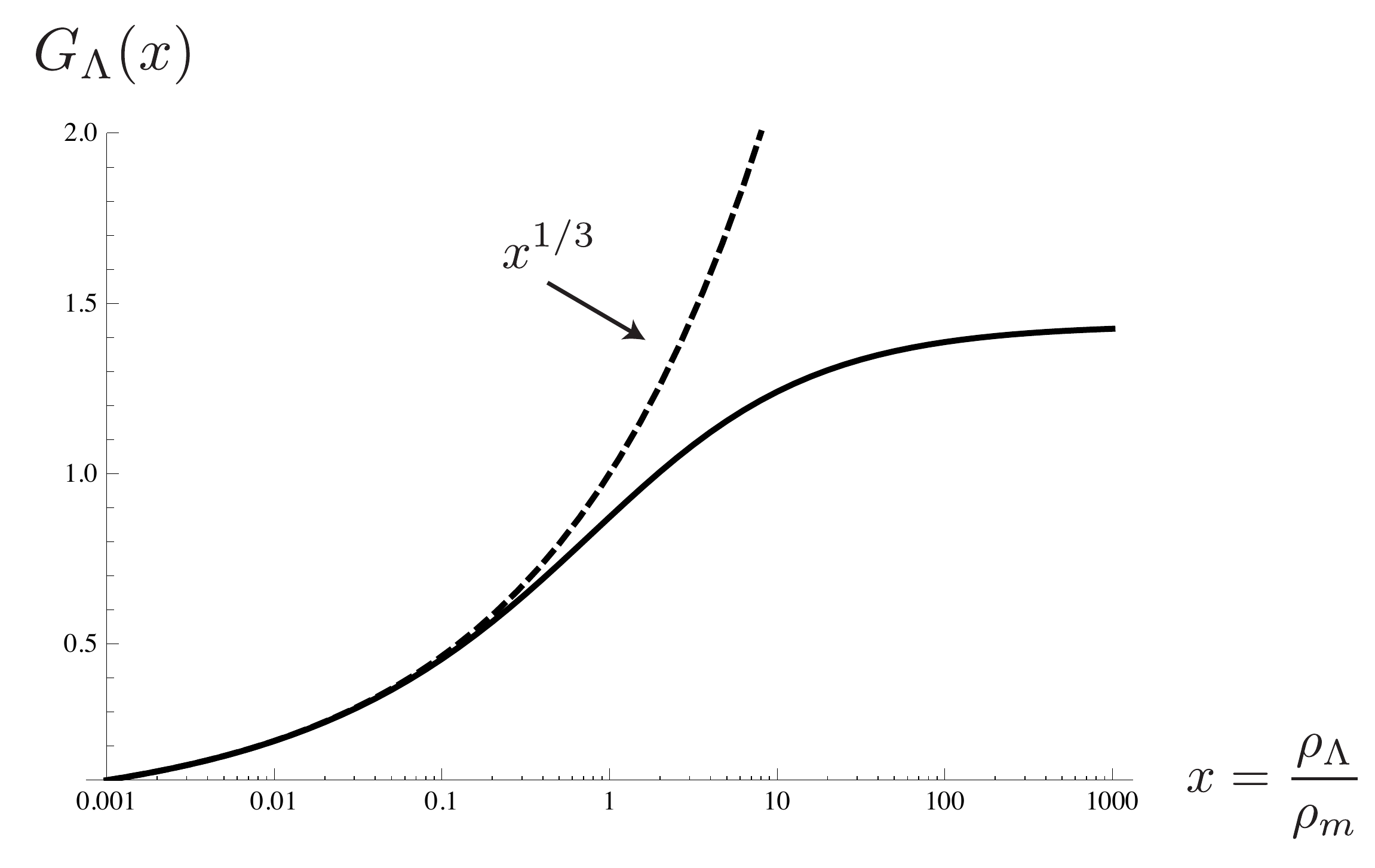}
\caption{The growth factor Eq.~\eqref{eq-gl} (solid line), which behaves like $x^{1/3}$ (dashed line) during the matter era, and asymptotes to a constant value well above $x(t_\Lambda)=1$.}
\label{fig-growth}
\end{figure}

\subsection{Late-Time Extrapolation of Numerical Results}
\label{sec-extrapolate}

Available Boltzmann codes do not offer output for negative redshifts. In order to estimate the smoothed density contrast $\sigma_R$ in this regime, we extrapolate our numerical results for $\sigma_R(z)$ from positive to negative $z$, i.e., from $a<1$ to $a>1$. The most straightforward approach would be a linear extrapolation in some time variable, fitting both the value and the derivative of $\sigma_R$ at $z=0$. However, there is a physical effect that we must incorporate analytically: vacuum domination turns off structure growth on all scales. This effect is not strong enough at $z=0$ to have a significant imprint on the value or time derivative of $\sigma_R$. However, the effect is also rather simple, and thus easy to incorporate analytically.

\begin{figure*}[t]
\centering
\includegraphics[width=6.5in]{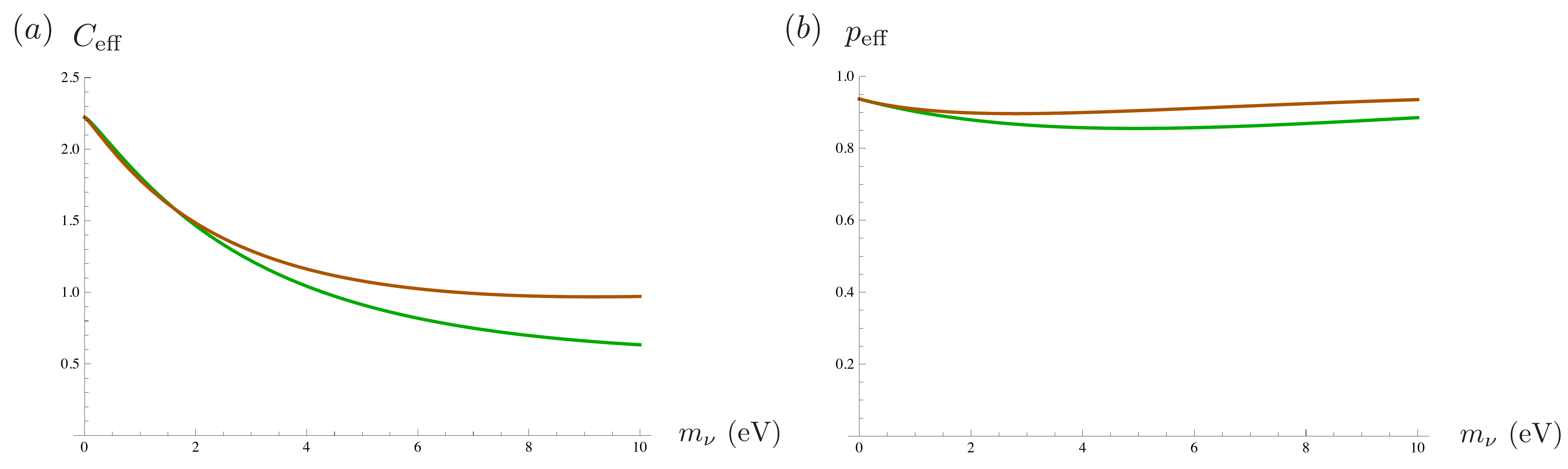}
\caption{The parameters $C_{\rm eff}$ and $p_{\rm eff}$, for normal (orange, top) and degenerate (green, bottom) hierarchies, obtained by fitting Eq.~(\ref{eq-sigmafit}) to CAMB output for $\sigma_R$ and its derivative at $z=0$. The resulting fitting function for $\sigma_R(z)$ is used to compute the Press-Schechter factor at negative redshift only. Note that  $p_{\rm eff}\approx 1$ throughout. This may seem surprising, but it is consistent with our earlier finding that at large neutrino masses, the scales whose power contributes dominantly to $\sigma_R$ are precisely the ones on which free-streaming is not effective. This is closely related to the discrepancy we find with Ref.~\cite{TegVil03}, whose estimate $p_{\rm eff}\approx p(k_{\rm gal})\approx 1-8f_\nu$ would yield a monotonically decreasing curve in (b).}
\label{fig-Ceffandp}
\end{figure*}

In a universe with negligible neutrino mass, the CDM density contrast grows as~\cite{MarSha97,Hea77} 
\begin{equation}
\delta\propto G_\Lambda(x)\equiv 
\frac{5}{6}\sqrt{1+\frac{1}{x}} \int_0^x \frac{dy}{y^{1/6}(1+y)^{3/2}}~,
\label{eq-gl}
\end{equation}
where
\begin{equation}
x\equiv \frac{\rho_\Lambda}{\rho_m} = \left. \frac{\Omega_\Lambda}{\Omega_m}\right|_{z=0} (1+z)^{-3}~.
\end{equation}
As seen in Fig.~\ref{fig-growth}, density perturbations grow like the scale factor during the matter dominated era; they asymptote to a constant value at times $t>t_\Lambda$. 

With nonzero neutrino mass, a reasonable approximation is obtained by combining the analytic result for the matter era, Eq.~(\ref{eq-deltap}), with the $m_\nu=0$ transition to the vacuum dominated era:
\begin{eqnarray} 
\delta & \propto & G_\Lambda(x)~~~\, (k< k_{\rm{nr}})~,\\
\delta & \propto & G_\Lambda(x)^p~~ (k> k_{\rm{nr}})~.
\end{eqnarray}
Recall that $P_{cc}(k)\propto \delta_c(k)^2$ by Eq.~(\ref{eq-pdelta}).

In order to improve on this result, we can incorporate the information gained from the use of Boltzmann codes. Instead of computing $p$ and $k_{\rm FS}$ analytically as described in the previous subsection, we can read off a slope $p(k)$ from the numerical output near $z=0$:
\begin{equation}
p(k) \equiv \frac{1}{2}\frac{d\log P_{cc}(x)}{d\log G_\Lambda(x)}~.
\end{equation}
We can also fix the constant of proportionality $C$ by matching the magnitude of $P_{cc}$ obtained from CAMB at $z=0$. This yields a semi-analytic power spectrum as a function of time, for any fixed $k$ and fixed neutrino mass:
\begin{equation}
P_{\rm cc}(x) = C G_\Lambda(x)^{2p(k)}~.
\end{equation}

In practice, it is cumbersome to extrapolate the power at each wave number only to integrate over scales to obtain the smoothed density contrast. By the late time corresponding to $z=0$, for any neutrino mass, we expect that the integral in Eq.~(\ref{eq-sp}) is dominated by the power at some scale $k$ and will remain dominated by the same scale in the future ($z<0$). For small neutrino masses, this scale will be set by the galaxy scale; for large $m_\nu$ it will be the scale of the peak of the spectrum $k^3 P(k)$. We incorporate this by matching the analytic growth for $z<0$ directly to the numerical results for $\sigma_R(x)$ at $z=0$.  For every $m_\nu$, we compute
\begin{eqnarray} 
p_{\rm eff} & \equiv & \left. \frac{d\log \sigma_R(x)}{d\log G_\Lambda(x)}\right|_{z=0}~,\\
C_{\rm eff} & \equiv & \left. \frac{\sigma_R}{G_\Lambda(x)^{p_{\rm eff}}}\right|_{z=0}~
\end{eqnarray} 
from the CAMB output for small nonnegative redshifts. The results are shown in Fig.~\ref{fig-Ceffandp}.

As our semi-analytic approximation entering the Press-Schechter factor $F$ for $z<0$ we use
\begin{equation}
\sigma_R(z) = C_{\rm eff} G_\Lambda(x(z))^{p_{\rm eff}}~~~~~\text{[used for}~z<0~\text{only]}
\label{eq-sigmafit}
\end{equation}
with $G_\Lambda$ given by Eq.~(\ref{eq-gl}). We have checked that the same formula provides an excellent fit to the numerical results at $z>0$, as one would expect. However, we stress again that we use the output from the CAMB code in this regime, not the fitting function. Moreover, the regime $z>0$ dominates in our calculation because the comoving volume of the causal patch decreases rapidly below $z=2$.

\section{Cooling and Galaxy Formation}
\label{sec-cool}

In this Appendix, we review the basic time scales that are believed to control cooling flows in dark matter halos. Our discussion closely follows Ref.~\cite{BouHal09}, where further details and references can be found.

Baryonic gas will fall into the gravitational well of newly formed dark matter halos. The baryons are thus shock-heated to high temperatures. In order for stars to form, the baryonic gas must cool and condense.  The initial temperature of the baryons is called the virial temperature. By the virial theorem,
\begin{equation}
\frac{G M_{\rm vir}\mu}{5 R_{\rm vir}} 
  = T_{\rm vir}~,
\label{eq:vir-theor}
\end{equation}
where $M_{\rm vir}$ is the mass of the halo and $R_{\rm vir}$ is its virial radius. In the regime of interest for us, $T_{\rm vir}$ is large enough to ionize hydrogen. Then one can take the average molecular mass $\mu$ to be $m_p/2$, where $m_p$ is the mass of the proton. With $M_{\rm vir}=\frac{4\pi}{3}\rho_{\rm vir} R_{\rm vir}^3$ one finds
\begin{equation}
  T_{\rm vir} \propto M_{\rm vir}^{2/3} \rho_{\rm vir}^{1/3}~,
\label{eq:T_vir}
\end{equation}
where the ``constants'' of proportionality depend negligibly on $M_{\rm vir}$. 

The timescale for cooling by bremsstrahlung is 
\begin{equation}
t_{\rm brems}\propto \frac{T_{\rm vir}^{1/2}}{\rho_{\rm vir}} 
\propto \frac{M_{\rm vir}^{1/3}}{\rho_{\rm vir}^{5/6}}~.
\end{equation}
We will be interested in how this timescale compares to the age of the universe when the halo virializes,
\begin{equation}
t_{\rm vir}\propto \rho_{\rm vir}^{-1/2}~.
\label{eq-rvtv}
\end{equation}

If $t_{\rm brems}\lesssim t_{\rm vir}$, then galaxy formation can be treated as instantaneous, i.e., as occurring nearly simultaneously with halo formation. Keeping track of all constants~\cite{BouHal09}, one finds that this case corresponds to  
\begin{equation}
M_{\rm vir} t_{\rm vir} ^2 \lesssim (10^{12} M_\odot)(2.2 \mbox{Gyr})^2~.
\label{eq-mt2}
\end{equation}
In the opposite case, $M_{\rm vir} t_{\rm vir} ^2 \gg (10^{12} M_\odot)(2.2 \mbox{Gyr})^2$, we have $t_{\rm brems} \gg t_{\rm vir}$. In halos with these mass and virialization time combinations, galaxy formation cannot be treated as instantaneous. Instead, it takes a much greater time $t_{\rm brems} \gg t_{\rm vir}$ to convert a comparable fraction of baryons into stars. (If feedback or major mergers disrupt the cooling flow, the contrast would be even more drastic, but we will not assume this here.)

The above analysis assumed cooling of unbound charged particles by bremsstrahlung. This approximation is best for virial temperatures above $10^7$\,K. At lower temperatures the cooling function is quite complicated, but one can get an estimate by treating it as independent of $T_{\rm vir}$ in some range~\cite{BouLei08}. With this approximation, one obtains that the cooling condition is satisfied for
\begin{equation}
M_{\rm vir}^2\, t_{\rm vir} < (10^{12} M_\odot)^2 (5.3\,\mbox{Gyr})~.
\label{eq-bl2}
\end{equation}
With either scaling, one finds again that cooling is inefficient if $M_{\rm vir} > 10^{12} M_\odot$, particularly for late virialization $t_{\rm vir} \gtrsim 10$\,Gyr.

\clubpenalty=0

So far, we have neglected the effects of the cosmological constant. For halos that form deep in the vacuum dominated era, one should use $\rho_{\rm vir} \sim\rho_\Lambda$ instead of Eq.~(\ref{eq-rvtv}). But such halos contribute negligibly in the causal patch because they will be exponentially dilute.

We have also neglected neutrinos. However, Eq.~(\ref{eq-mt2}) is sufficiently general to capture their main effect, which is to change the relation between $M_{\rm vir}$ and $t_{\rm vir}$. In a universe with $m_\nu\ll 8$\,eV, $t_{\rm vir}$ grows logarithmically with $M_{\rm vir}$ for overdensities of a fixed relative amplitude. For $10^{12} M_\odot$ halos forming from $1\sigma$ ($2\sigma$) overdensities, $t_{\rm vir}\approx 3.6$\,Gyr ($t_{\rm vir}\approx 1.3$\,Gyr) and by Eq.~(\ref{eq-mt2}), cooling fails (succeeds).

In a universe with $m_\nu\gtrsim 8$\,eV, however, small scale power is so suppressed that structure formation proceeds in a top-down manner. (This is shown in detail in the main text.) Then structure on all scales forms much later than $2.4$\,Gyr.  Moreover, smaller structure is embedded in larger halos, which set the virial mass that enters Eq.~(\ref{eq-mt2}). Hence, the timescale for a significant fraction of baryons to form stars is at least $t_{\rm brems} \gg t_{\rm vir} \gg O(\mbox{Gyr})$.

\bibliographystyle{utcaps}
\bibliography{allneutrinos}
\end{document}